\pdfoutput=1
\documentclass[journal,10pt]{IEEEtran}
\usepackage{cite}

\ifCLASSINFOpdf
\else
\fi
\usepackage[cmex10]{amsmath}
\usepackage{algorithmic}

\usepackage{array}

\usepackage[tight,footnotesize]{subfigure}
\usepackage{color}

\usepackage{amssymb}
\usepackage{amsthm}
\theoremstyle{plain}
\newtheorem{thm}{Theorem}
\newtheorem{lem}[thm]{Lemma}

\newtheorem{claim}{Claim}
\theoremstyle{definition}
\newtheorem{defn}{Definition}
\theoremstyle{remark}

\input xy
\xyoption{all}
\xyoption{color}

\usepackage{wasysym,multirow,pifont,hyperref,algorithm2e}
\usepackage{xcolor}
\usepackage{graphicx}
\usepackage{stfloats}

\IEEEiedlistdecl

\begin{document}
%
\title{On the Dynamics of the Error Floor Behavior\\ in (Regular) LDPC Codes}
%
%
%

\author{Christian~Schlegel,~\IEEEmembership{Senior Member,~IEEE,}
        Shuai~Zhang
\thanks{C.~Schlegel and S.~Zhang are with the High Capacity Digital Communications
Laboratory (HCDC), Electrical and Computer Engineering Department,
University of Alberta, Edmonton AB, T6G 2V4, CANADA (e-mail: \{schlegel, szhang4\}@ece.ualberta.ca).}
}

%
%

\markboth{Submitted to IEEE Transactions on Information Theory in February, 2009}
{Schlegel \MakeLowercase{\textit{et al.}}: On the Dynamics of the
Error Floor Behavior in (Regular) LDPC Codes}
%

\IEEEpubid{0000--0000/00\$00.00~\copyright~2009 IEEE}
\IEEEpubidadjcol


\maketitle

\begin{abstract}
It is shown that dominant trapping sets of regular LDPC codes, so
called absorption sets, undergo a two-phased dynamic behavior in the
iterative message-passing decoding algorithm. Using a linear dynamic
model for the iteration behavior of these sets, it is shown that
they undergo an initial geometric growth phase which stabilizes in a
final bit-flipping behavior where the algorithm reaches a fixed
point. This analysis is shown to lead to very accurate numerical
calculations of the error floor bit error rates down to error rates
that are inaccessible by simulation. The topology of the dominant
absorption sets of an example code, the IEEE 802.3an $(2048,1723)$
regular LDPC code, are identified and tabulated using topological
relationships in combination with search algorithms.
\end{abstract}


%
\IEEEpeerreviewmaketitle

\section{Summary}
%
%
%
%
\IEEEPARstart{T}{he} error floor in modern graph-based error control
codes such as low-density parity-check codes is caused by inherent
structural weaknesses in the code's interconnect network. The
iterative message passing algorithm cannot overcome these weaknesses
and gets trapped in error patterns which are easily identifiable as
erroneous (in LDPC codes), and are thus not valid codewords, but
difficult to overcome or correct \cite{perez08, perez09}. These weaknesses were termed {\em
trapping sets} by Richardson in \cite{Rich04}, a summary definition
for the patterns on which the message passing algorithm fails for
Gaussian channels. These trapping sets are dependent on the code,
the channel used, and to a lesser degree also on the details of the
decoding algorithm. Prior work in identifying the weaknesses of LDPC
codes on erasure channels led to the definition of {\em stopping
sets} in \cite{ChanProiTelaRichUrba02}. Stopping sets, being the
weaknesses of LDPC codes on erasure channels, also play a role on
Gaussian channels, but are not typically the dominant error
mechanisms. In \cite{Zhanetal06} the authors define {\em absorption
sets}, which are the subgraphs of the code graph on which the
Gallager bit-flipping decoding algorithms fail for binary symmetric
channels. The authors observed that these absorption sets also show
up as the dominant trapping sets in certain structured LDPC codes.
In \cite{Zhanetal08} they devise post-processing methods to reduce
the effects of these absorption sets and lower the error floor of
the codes in question.

In this paper we present a linear algebraic approach to the dynamic
behavior of absorption sets. We show that these sets follow a
geometric growth phase during early iterations where messages inside
the absorption set grow towards a largest eigenvector which characterizes
the absorption set. The seemingly erratic behavior
of the messages at early iterations is due to the decreasing
influence of lesser eigenvectors. We define the {\em gain} of an
absorption set and show how it affects the influence of the
extrinsic messages that flow into the absorption set at each
iteration from the remainder of the code network. The importance of
set extrinsic information was already informally observed in
\cite{YanRyaLi04}, who reported a lowering of the error floor with
increased extrinsic connectivity. We use our analysis to produce
accurate error formulas for the error floor BER/FER and support
these results with importance sampling simulations targeting the
absorption sets.

As illustration we carefully identify and classify absorption sets
of the regular $(2048,1723)$ LDPC code recently designed in
\cite{DjuXuAbdLin03}, which is used in the IEEE 802.3an
standard.

Topological features of dominant absorption sets are identified and a search algorithm
is presented which finds the leading dominant sets.


\section{Background}

\IEEEpubidadjcol

Stopping sets completely determine the performance of graph-based
decoding of LDPC codes on erasure channels, i.e., on channels where
the transmitted binary symbols are either received correctly, or are
erased. A complete statistical treatment of stopping sets was given
in \cite{ChanProiTelaRichUrba02}. Aptly named, a stopping set is a
subset of uncorrected variable nodes where the decoder stops, i.e.,
makes no further correction progress. It is simply
defined as:

\begin{defn}
A stopping set $\cal S$ is a set of variable nodes, all of whose
neighboring check nodes are connected to the set $\cal S$ at least twice.
\end{defn}

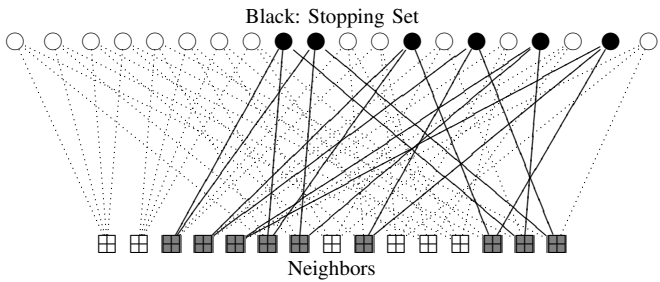
\begin{figure}[!t]
\centering
\setlength{\unitlength}{0.8mm}
\begin{picture}(130,44)
\setlength{\fboxsep}{0mm} \setlength{\fboxrule}{0mm}
\put(0,37){$$\xymatrix@M=0pt@W=0pt@R=70pt@C=-1.7pt {
\Circle\ar@{.}[drrrrr]\ar@{.}[drrrrrrrrrrrrrrr]\ar@{.}[drrrrrrrrrrrrrrrrrrrrrrrrr]&& \Circle\ar@{.}[drrr]\ar@{.}[drrrrrrrrrrrrrrr]\ar@{.}[drrrrrrrrrrrrrrrrrrrrrrrrr]&&\Circle\ar@{.}[dr]\ar@{.}[drrrrrrrrrrrrrrr]\ar@{.}[drrrrrrrrrrrrrrrrrrrrrrrrr]&& \Circle\ar@{.}[dl]\ar@{.}[drrrrrrrrrrrrrrr]\ar@{.}[drrrrrrrrrrrrrrrrrrrrrrrrr]&& \Circle\ar@{.}[dl]\ar@{.}[drrrrrrr]\ar@{.}[drrrrrrrrrrrrrrrrrrrrrrrrr]&& \Circle\ar@{.}[dlll]\ar@{.}[drrrrrrr]\ar@{.}[drrrrrrrrrrrrrrr]&& \Circle\ar@{.}[dlllll]\ar@{.}[drrrrrrr]\ar@{.}[drrrrrrrrrrrrrrr]&& \Circle\ar@{.}[dlllllll]\ar@{.}[drrrrrrrrr]\ar@{.}[drrrrrrrrrrrrrrr]&& \CIRCLE\ar@{-}[dlllllll]\ar@{-}[dl]\ar@{-}[drrrrrrrrrrrrrrr]&& \CIRCLE\ar@{-}[dlllllllll]\ar@{-}[dl]\ar@{-}[drrrrrrrrrrrrrrr]&& \Circle\ar@{.}[dlllllllllll]\ar@{.}[dr]\ar@{.}[drrrrrrr]&& \Circle\ar@{.}[dlllllllllllll]\ar@{.}[dr]\ar@{.}[drrr]&& \CIRCLE\ar@{-}[dlllllllllllll]\ar@{-}[dlllllllll]\ar@{-}[drrrrr]&& \Circle\ar@{.}[dlllllllllllllll]\ar@{.}[dlllllll]\ar@{.}[drrrrr]&& \CIRCLE\ar@{-}[dlllllllllllllllll]\ar@{-}[dlllllll]\ar@{-}[drrrrr]&& \Circle\ar@{.}[dlllllllllllllllllll]\ar@{.}[dlllllll]\ar@{.}[dlll]&& \CIRCLE\ar@{-}[dlllllllllllllllllll]\ar@{-}[dlllllllllllllll]\ar@{-}[dl]&& \Circle\ar@{.}[dlllllllllllllllllllll]\ar@{.}[dlllllllllllllll]\ar@{.}[dlllllllll]&& \CIRCLE\ar@{-}[dlllllllllllllllllllllll]\ar@{-}[dlllllllllllllll]\ar@{-}[dlllllll]&& \Circle\ar@{.}[dlllllllllllllllllllllllll]\ar@{.}[dlllllllllllllll]\ar@{.}[dlllll]\\
&\hspace{3.5mm}& &\hspace{3.5mm}& &\boxplus& & \boxplus & & \colorbox{gray}{$\boxplus$}& & \colorbox{gray}{$\boxplus$}& & \colorbox{gray}{$\boxplus$}& & \colorbox{gray}{$\boxplus$}& & \colorbox{gray}{$\boxplus$}& & \boxplus& & \colorbox{gray}{$\boxplus$}& & \boxplus& & \boxplus& & \boxplus& & \colorbox{gray}{$\boxplus$}& & \colorbox{gray}{$\boxplus$}& & \colorbox{gray}{$\boxplus$}& &\hspace{3.5mm}& &\hspace{3.5mm}& \\
}$$} \put(40,40){\footnotesize Black: Stopping Set} \put(47,-2){\footnotesize Neighbors}
\end{picture}
\caption{Example of a stopping set.} \label{fig:stopping}
\end{figure}

\figurename \ref{fig:stopping} shows an example of a stopping set.
It is quite straightforward to see that if erasure decoding is
performed following Gallager's decoding algorithm \cite{SchPer04} the
variable values in the stopping set cannot be reconstructed. Valid
codewords are trivially stopping sets, but the set of stopping sets
is larger than the set of valid codewords.

An absorption set is an extension of the notion of a stopping set to
the binary-symmetric channels \cite{Zhanetal06,Zhanetal08}, and is
defined as:

\begin{defn}\label{def1}
An absorption set $\cal A$ is a set of variable nodes, such that the
majority of each variable node's neighbors are connected to the set $\cal A$
an even number of times.
\end{defn}

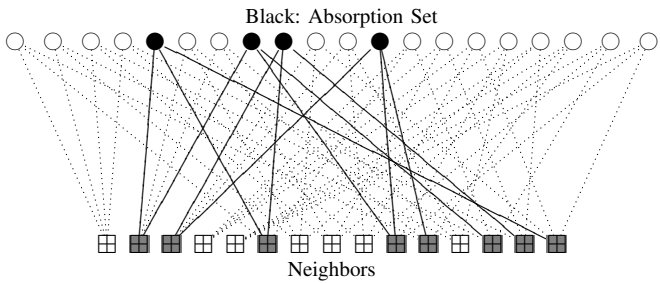
\begin{figure}[!t]
\centering
\setlength{\unitlength}{0.8mm}
\begin{picture}(130,44)
\setlength{\fboxsep}{0mm} \setlength{\fboxrule}{0mm}
\put(0,37){$$\xymatrix@M=0pt@W=0pt@R=70pt@C=-1.7pt {
\Circle\ar@{.}[drrrrr]\ar@{.}[drrrrrrrrrrrrrrr]\ar@{.}[drrrrrrrrrrrrrrrrrrrrrrrrr]&& \Circle\ar@{.}[drrr]\ar@{.}[drrrrrrrrrrrrrrr]\ar@{.}[drrrrrrrrrrrrrrrrrrrrrrrrr]&&\Circle\ar@{.}[dr]\ar@{.}[drrrrrrrrrrrrrrr]\ar@{.}[drrrrrrrrrrrrrrrrrrrrrrrrr]&& \Circle\ar@{.}[dl]\ar@{.}[drrrrrrrrrrrrrrr]\ar@{.}[drrrrrrrrrrrrrrrrrrrrrrrrr]&& \CIRCLE\ar@{-}[dl]\ar@{-}[drrrrrrr]\ar@{-}[drrrrrrrrrrrrrrrrrrrrrrrrr]&& \Circle\ar@{.}[dlll]\ar@{.}[drrrrrrr]\ar@{.}[drrrrrrrrrrrrrrr]&& \Circle\ar@{.}[dlllll]\ar@{.}[drrrrrrr]\ar@{.}[drrrrrrrrrrrrrrr]&& \CIRCLE\ar@{-}[dlllllll]\ar@{-}[drrrrrrrrr]\ar@{-}[drrrrrrrrrrrrrrr]&& \CIRCLE\ar@{-}[dlllllll]\ar@{-}[dl]\ar@{-}[drrrrrrrrrrrrrrr]&& \Circle\ar@{.}[dlllllllll]\ar@{.}[dl]\ar@{.}[drrrrrrrrrrrrrrr]&& \Circle\ar@{.}[dlllllllllll]\ar@{.}[dr]\ar@{.}[drrrrrrr]&& \CIRCLE\ar@{-}[dlllllllllllll]\ar@{-}[dr]\ar@{-}[drrr]&& \Circle\ar@{.}[dlllllllllllll]\ar@{.}[dlllllllll]\ar@{.}[drrrrr]&& \Circle\ar@{.}[dlllllllllllllll]\ar@{.}[dlllllll]\ar@{.}[drrrrr]&& \Circle\ar@{.}[dlllllllllllllllll]\ar@{.}[dlllllll]\ar@{.}[drrrrr]&& \Circle\ar@{.}[dlllllllllllllllllll]\ar@{.}[dlllllll]\ar@{.}[dlll]&& \Circle\ar@{.}[dlllllllllllllllllll]\ar@{.}[dlllllllllllllll]\ar@{.}[dl]&& \Circle\ar@{.}[dlllllllllllllllllllll]\ar@{.}[dlllllllllllllll]\ar@{.}[dlllllllll]&& \Circle\ar@{.}[dlllllllllllllllllllllll]\ar@{.}[dlllllllllllllll]\ar@{.}[dlllllll]&& \Circle\ar@{.}[dlllllllllllllllllllllllll]\ar@{.}[dlllllllllllllll]\ar@{.}[dlllll]\\
&\hspace{3.5mm}& &\hspace{3.5mm}& &\boxplus& &\colorbox{gray}{$\boxplus$} & & \colorbox{gray}{$\boxplus$}& & \boxplus& & \boxplus& & \colorbox{gray}{$\boxplus$}& & \boxplus& & \boxplus& & \boxplus& & \colorbox{gray}{$\boxplus$}& & \colorbox{gray}{$\boxplus$}& & \boxplus& & \colorbox{gray}{$\boxplus$}& & \colorbox{gray}{$\boxplus$}& & \colorbox{gray}{$\boxplus$}& &\hspace{3.5mm}& &\hspace{3.5mm}& \\
}$$} \put(40,40){\footnotesize Black: Absorption Set} \put(47,-2){\footnotesize Neighbors}
\end{picture}
\caption{Example of an absorption set.} \label{fig:absorption}
\end{figure}

\figurename \ref{fig:absorption} shows an example absorption set. It
can be verified that Gallager-type bit flipping decoding will not be
able to correct an absorption  set, since a majority of messages
impinging on each variable node will retain the erroneous sign for
each iteration. Consequently, the algorithm locks up.

\section{LDPC Codes on the Gaussian Channel}

The Gaussian channel is different from the binary symmetric and
binary erasure channels and causes a more complicated error behavior
on LDPCs. Richardson \cite{Rich04} first seriously explored the
error floor of LDPCs on Gaussian channels and defined {\em trapping
sets} as the failure mechanism. Noting that typically very few
trapping sets dominate the error floor region he proposed a
semi-analytical method which amounts to a variant of importance
sampling to numerically predict the error floor from the knowledge
of a code's trapping sets.

While finding trapping sets remained a largely open problem,
\cite{Zhanetal06} observed that in certain structured LDPCs the
dominant trapping sets are absorption sets, i.e., the failure
mechanism of the code on binary symmetric channels. In
\cite{Zhanetal08}, algorithmic modifications were proposed to
``eliminate'' the error floor caused by these absorption sets.

Due to its popularity and extensive exposure we will
concentrate on the $(2048,1723)$ regular LDPC code
\cite{DjuXuAbdLin03} used in the IEEE 802.3an standard. This code
has been extensively analyzed. It has a low error floor that appears
at $E_b/N_0\approx 5$dB at a BER of $10^{-12}$, that is too low to
be efficiently explored using conventional simulations\footnote{Even
an FPGA-based simulation running at 100Mb/s requires about a week
for a single data point.}.

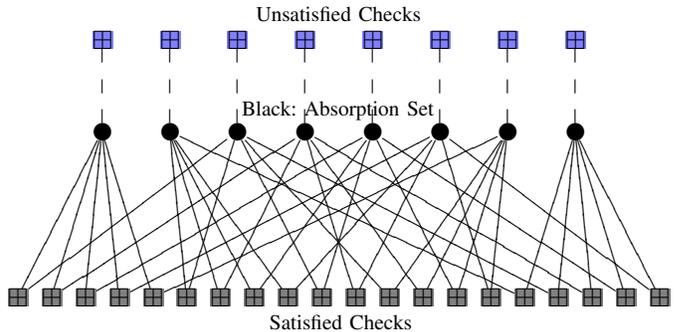
\begin{figure}[!t]
\centering
\setlength{\unitlength}{1mm}
\begin{picture}(170,42)
\setlength{\fboxsep}{0mm} \setlength{\fboxrule}{0mm}
\put(0,36.5){$$\xymatrix@M=0pt@W=0pt@R=28pt@C=-2pt {
&&&&& \colorbox{blue!50}{$\boxplus$}\ar@{--}[d]&&&& \colorbox{blue!50}{$\boxplus$}\ar@{--}[d]&&&& \colorbox{blue!50}{$\boxplus$}\ar@{--}[d]&&&& \colorbox{blue!50}{$\boxplus$}\ar@{--}[d]&&&& \colorbox{blue!50}{$\boxplus$}\ar@{--}[d]&&&& \colorbox{blue!50}{$\boxplus$}\ar@{--}[d]&&&& \colorbox{blue!50}{$\boxplus$}\ar@{--}[d]&&&& \colorbox{blue!50}{$\boxplus$}\ar@{--}[d]&&&&& \\
&&&&& \CIRCLE\ar@{-}[ddlllll]\ar@{-}[ddlll]\ar@{-}[ddl]\ar@{-}[ddr]\ar@{-}[ddrrr]&&&& \CIRCLE\ar@{-}[ddr]\ar@{-}[ddrrr]\ar@{-}[ddrrrrr]\ar@{-}[ddrrrrrrr]\ar@{-}[ddrrrrrrrrrrrrrrrrrrrrr]&&&& \CIRCLE\ar@{-}[ddlllllllllllll]\ar@{-}[ddlll]\ar@{-}[ddrrrrr]\ar@{-}[ddrrrrrrrrrrrrrrrrrrr]\ar@{-}[ddrrrrrrrrr]&&&& \CIRCLE\ar@{-}[ddlllllllllllllll]\ar@{-}[ddlllll]\ar@{-}[ddrrr]\ar@{-}[ddrrrrrrrrrrrrrrrrr]\ar@{-}[ddrrrrrrr]&&&& \CIRCLE\ar@{-}[ddlll]\ar@{-}[ddlllllll]\ar@{-}[ddlllllllllllllllll]\ar@{-}[ddrrrrrrrrrrrrrrr]\ar@{-}[ddrrrrr]&&&& \CIRCLE\ar@{-}[ddrrrrrrrrrrrrr]\ar@{-}[ddrrr]\ar@{-}[ddlllll]\ar@{-}[ddlllllllllllllllllll]\ar@{-}[ddlllllllll]&&&& \CIRCLE\ar@{-}[ddl]\ar@{-}[ddlll]\ar@{-}[ddlllll]\ar@{-}[ddlllllll]\ar@{-}[ddlllllllllllllllllllll]&&&& \CIRCLE\ar@{-}[ddlll]\ar@{-}[ddl]\ar@{-}[ddr]\ar@{-}[ddrrr]\ar@{-}[ddrrrrr]&&&&& \\
&&&&&&&&&&&&&&&&&&&&&&&&&&&&&&&&&&&&&&\\
\colorbox{gray}{$\boxplus$}&\hspace{9pt}&
\colorbox{gray}{$\boxplus$}&\hspace{9pt}&
\colorbox{gray}{$\boxplus$}&\hspace{9pt}&
\colorbox{gray}{$\boxplus$}&\hspace{9pt}&
\colorbox{gray}{$\boxplus$}&\hspace{9pt}&
\colorbox{gray}{$\boxplus$}&\hspace{9pt}&
\colorbox{gray}{$\boxplus$}&\hspace{9pt}&
\colorbox{gray}{$\boxplus$}&\hspace{9pt}&
\colorbox{gray}{$\boxplus$}&\hspace{9pt}&
\colorbox{gray}{$\boxplus$}&\hspace{9pt}&
\colorbox{gray}{$\boxplus$}&\hspace{9pt}&
\colorbox{gray}{$\boxplus$}&\hspace{9pt}&
\colorbox{gray}{$\boxplus$}&\hspace{9pt}&
\colorbox{gray}{$\boxplus$}&\hspace{9pt}&
\colorbox{gray}{$\boxplus$}&\hspace{9pt}&
\colorbox{gray}{$\boxplus$}&\hspace{9pt}&
\colorbox{gray}{$\boxplus$}&\hspace{9pt}&
\colorbox{gray}{$\boxplus$}&\hspace{9pt}&
\colorbox{gray}{$\boxplus$}&\hspace{9pt}&
\colorbox{gray}{$\boxplus$} }$$} \put(31.1,26.5){\footnotesize Black: Absorption
Set} \put(34.8,-2){\footnotesize Satisfied Checks} \put(33,39){\footnotesize Unsatisfied
Checks}
\end{picture}
\caption{A dominant absorption set of the IEEE 802.3an code (not all
check node connections shown).} \label{fig:trapping}
\end{figure}

\figurename \ref{fig:trapping} shows the structure of the {\em
dominant} absorption set of this code (see also \cite[\figurename
2]{Zhanetal08}). There are $14,272$ such sets in the $(2048,1723)$
code of \cite{DjuXuAbdLin03}. They dominate the error floor since
they are the minimal absorption sets in this code (for definition of
minimal and dominant, see Definition \ref{defn4}).

\subsection{Finding Dominant Absorption Sets}

The $(2048,1723)$ rate $0.8413$ regular LDPC code
\cite{DjuXuAbdLin03} considered here has a structured parity-check
matrix:
$$\mathbf{H}=\begin{bmatrix}
\mathbf{\sigma}_{11} &\mathbf{\sigma}_{12} &\mathbf{\sigma}_{13} &\cdots &\mathbf{\sigma}_{1,32} \\
\mathbf{\sigma}_{21} &\mathbf{\sigma}_{22} &\mathbf{\sigma}_{23} &\cdots &\mathbf{\sigma}_{2,32} \\
\mathbf{\sigma}_{31} &\mathbf{\sigma}_{32} &\mathbf{\sigma}_{33} &\cdots &\mathbf{\sigma}_{3,32} \\
\vdots &\vdots &\vdots &\ddots &\vdots \\
\mathbf{\sigma}_{61} &\mathbf{\sigma}_{62} &\mathbf{\sigma}_{63} &\cdots &\mathbf{\sigma}_{6,32} \\
\end{bmatrix}_{384\times 2048}$$ where each $\mathbf{\sigma}_{ij}$ is a $64\times
64$ permutation matrix.

\begin{defn}
Let $(a,b)$ denote an absorption set, where $a$ is the size of the
set (number of variable nodes) and $b$ is the extrinsic message
degree (EMD), i.e., the cardinality of the set of the neighboring
check nodes that are connected to the set an odd number of times (the unsatisfied checks).
\end{defn}

For example, \figurename \ref{fig:absorption} and \figurename
\ref{fig:trapping} show $(4,4)$ and $(8,8)$ absorption sets,
respectively.

Table \ref{table1} shows the first few absorption sets of the code in
\cite{DjuXuAbdLin03}.

\begin{table}[!t]
\renewcommand{\arraystretch}{1.1}
\caption{Absorption sets of IEEE 802.3an LDPC code.} \label{table1}
\centering
\begin{tabular}{|c|c|c|c|c|}\hline \bfseries $a$ & \bfseries $b$ & \bfseries Existence
& \bfseries Multiplicity & \bfseries Gain\\\hline\hline $<5$ & & No
&&\\\hline $5$ & $10$ & No && \\\hline \multirow{4}{*}{$6$} & $6$ &
\multirow{4}{*}{No} &&\\\cline{2-2}
 & $8$ & &&\\\cline{2-2}
 & $10$ & &&\\\cline{2-2}
 & $12$ & &&\\ \hline
 \multirow{8}{*}{$7$} & $0$ & \multirow{6}{*}{No}&&\\\cline{2-2}
 & $2$ &&& \\\cline{2-2}
 & $4$ &&& \\\cline{2-2}
 & $6$ &&& \\\cline{2-2}
 & $8$ &&& \\\cline{2-2}
 & $10$ &&& \\\cline{2-5}
 & $12$ & Yes & $65,472$\footnotemark & $3.29$\\ \cline{2-5}
 & $14$ & Yes & ? & $3$ \\ \hline
 \multirow{9}{*}{$8$} & $0$ & \multirow{4}{*}{No}&&\\\cline{2-2}
 & $2$ &&& \\\cline{2-2}
 & $4$ &&& \\\cline{2-2}
 & $6$ &&& \\\cline{2-5}
 & $8$ & Yes & $14,272$ & $4$\\\cline{2-5}
 & $10$ & No &&\\\cline{2-5}
 & $12$ & Yes & $44,416$ &$3.5$\\\cline{2-5}
 & $14$ & \multirow{2}{*}{Yes} & \multirow{2}{*}{?} &$3.25$\\\cline{2-2}\cline{5-5}
 & $16$ &  &&$3$\\ \hline
\multirow{7}{*}{$9$} & $0$ & \multirow{2}{*}{No} &&\\ \cline{2-2}
  & $2$ &  &&\\ \cline{2-5}
  & $4\leq b\leq10$ & ? &&\\ \cline{2-5}
  & $12$ & \multirow{4}{*}{Yes} & \multirow{4}{*}{?} &$3.67$\\ \cline{2-2}\cline{5-5}
  & $14$ & & &$3.44$\\ \cline{2-2}\cline{5-5}
  & $16$ &  & &$3.22$\\ \cline{2-2}\cline{5-5}
  & $18$ &  & &$3$\\ \hline
 \multirow{7}{*}{$10$} & $\leq8$ & ? &&\\ \cline{2-5}
   & $10$ & Yes & $>192$ & $4$\\ \cline{2-5}
 & $12$ & \multirow{5}{*}{Yes} & \multirow{5}{*}{?} &$3.8$\\\cline{2-2}\cline{5-5}
 & $14$ &  &  &$3.6$\\\cline{2-2}\cline{5-5}
 & $16$ & &&$3.4$\\\cline{2-2}\cline{5-5}
 & $18$ &  &&$3.2$\\\cline{2-2}\cline{5-5}
 & $20$ &  &&$3$\\ \hline
\end{tabular}
\end{table}

\begin{defn}\label{defn4}
(i) Let the ratio $b/a$ denote the average EMD for an $(a,b)$ absorption set.
(ii) An $(a,b)$ absorption set is called minimal if no $(a',b')$ absorption set exists with $a'<a$ and $b'/a' \leq b/a$, i.e., less variable nodes and smaller average EMD.
(iii) A minimal $(a,b)$ absorption set is called dominant if no $(a,b')$ absorption set exists with $b'<b$, i.e., smaller EMD.
\end{defn}

The smaller the absorption set, the more severe the effect on the
error floor. Thus our target is to find the dominant absorption sets
in terms of $a$, $b$ and $b/a$. Since the variable node degree
$d_v=6$, $a\geq5$ by the definition of absorption sets and the code
is $4$-cycle free. In addition, $b\in[0,2a]$ and must be even. So
let us start with $a=5$ to develop the numbers in Table
\ref{table1}. The coefficient $5-b/a$ is the gain of the absorption
set, which determines how fast the extrinsic information enters the
set --- see later.

\subsubsection{$a=5$}

\begin{figure}[!t]
\centering \subfigure[Check nodes shown.] 
{
    \label{fig1:sub:a}
$\xymatrix@M=0pt@W=0pt@R=20pt@C=-1pt
{
\color{blue}\boxplus&\hspace{9pt}&\color{blue}\boxplus&\hspace{9pt}&\color{blue}\boxplus&\hspace{9pt}&\color{blue}\boxplus&\hspace{9pt}&\color{blue}\boxplus&\hspace{9pt}&\color{blue}\boxplus&\hspace{9pt}&\color{blue}\boxplus&\hspace{9pt}&\color{blue}\boxplus&\hspace{9pt}&\color{blue}\boxplus&\hspace{9pt}&\color{blue}\boxplus\\
&\CIRCLE\ar@{--}[ul]\ar@{--}[ur]\ar@{-}[ddl]\ar@{-}[ddr]\ar@{-}[ddrrr]\ar@{-}[ddrrrrr]&&&&\CIRCLE\ar@{--}[ul]\ar@{--}[ur]\ar@{-}[ddlllll]\ar@{-}[ddrrr]\ar@{-}[ddrrrrr]\ar@{-}[ddrrrrrrr]&&&&\CIRCLE\ar@{--}[ul]\ar@{--}[ur]\ar@{-}[ddlllllll]\ar@{-}[ddl]\ar@{-}[ddrrrrr]\ar@{-}[ddrrrrrrr]&&&&\CIRCLE\ar@{--}[ul]\ar@{--}[ur]\ar@{-}[ddlllllllll]\ar@{-}[ddlll]\ar@{-}[ddr]\ar@{-}[ddrrrrr]&&&&\CIRCLE\ar@{--}[ul]\ar@{--}[ur]\ar@{-}[ddlllllllllll]\ar@{-}[ddlllll]\ar@{-}[ddl]\ar@{-}[ddr]&\\
&&&&&&&&&&&&&&&&&&\\
\boxplus&&\boxplus&&\boxplus&&\boxplus&&\boxplus&&\boxplus&&\boxplus&&\boxplus&&\boxplus&&\boxplus
 }$}\hfill
\subfigure[Check nodes hidden.] 
{
    \label{fig1:sub:b}
$\xymatrix@M=0pt@W=0pt@R=30pt@C=15pt
{
&&\CIRCLE\ar@{-}[dll]\ar@{-}[drr]\ar@{-}[ddl]\ar@{-}[ddr]&&\\\CIRCLE\ar@{-}[rrrr]\ar@{-}[dr]\ar@{-}[drrr]&&&&\CIRCLE\ar@{-}[dl]\ar@{-}[dlll]\\&\CIRCLE\ar@{-}[rr]&&\CIRCLE&
}$} \caption{The only possible topology of $(5,10)$ absorption
sets.}
\label{fig1:sub} 
\end{figure}
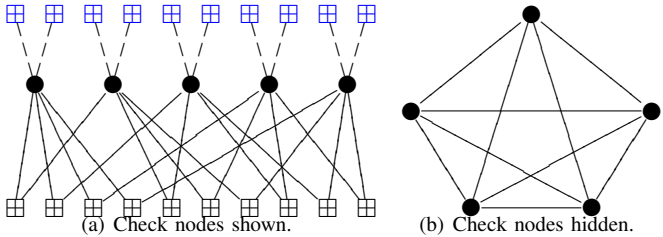

Clearly $b$ can only equal $10$ and there is only one possible
connecting topology shown in \figurename \ref{fig1:sub}.

\footnotetext{Only sets not contained in $(8,8)$ absorption sets are
counted --- see later.}

\begin{lem}\label{lema=5}
There are no size-$5$ absorption sets.
\end{lem}
\begin{IEEEproof} See Appendix \ref{proofa=5}.
\end{IEEEproof}

\subsubsection{$a=6$}

Let us introduce additional notations needed to prove the existence
of absorption sets.

\begin{defn}\label{constraints}
(i) For any variable node $v$ in an absorption set, let $\mathrm{Deg}(v)$
denote the number of neighboring check nodes of $v$ that are
connected to the set an even number of times. $\mathrm{Deg}(v)$ is the degree
of vertex $v$ in the topology graph with check nodes hidden. (ii)
Let an unordered array $[\mathrm{Deg}(v_i) : i=1,2,\ldots,a]$ denote a class
of $(a,b)$ absorption sets, where $\mathrm{Deg}(v_i)\in \{4,5,6\}$ and
$\sum\limits_{i=1}^a \mathrm{Deg}(v_i)=6a-b$.
\end{defn}

It is difficult to find absorption sets, even by making use of
Algorithm \ref{alg:mine}-type methods (see Appendix \ref{proofa=5}),
due to their extremely low appearance. Hence we need Definition
\ref{constraints} to classify the absorption sets first. For each
pair $(a,b)$, there may be several classes of absorption sets, and
each class may exhibit several topologies. What we are trying to do
is to reduce one unknown absorption set to a smaller absorption set
whose non-existence is known by eliminating nodes from the original
set. We can then argue that there is only a limited number of
topologies that have to be searched algorithmically.

\begin{thm}\label{thma=6}
There are no size-$6$ absorption sets.
\end{thm}
\begin{IEEEproof} See Appendix \ref{proofa=6}.
\end{IEEEproof}

\subsubsection{$a=7$}

\begin{thm}\label{thma=7}
There are no $(7,b)$ absorption sets with $b<12$. $(7,12)$
and $(7,14)$ absorption sets do exist.
\end{thm}
\begin{IEEEproof} See Appendix \ref{proofa=7}.
\end{IEEEproof}

\subsubsection{$a=8$}

First we show
\begin{lem}\label{lema=8}
For $b<8$, there exist no $(8,b)$ absorption sets.\footnote{As a corollary, since there is no $(8,0)$ absorption set, the minimum distance bound of this LDPC code \cite{DjuXuAbdLin03} is strengthened
to $d_{\mathrm{min}}\geq 10$. Therefore, there are no $(9,0)$ absorption sets since a $(9,0)$ absorption set is a length-$9$ codeword.} For $b=8$, there
exists no $(8,8)$ absorption set that contains a degree-$6$ variable node.
\end{lem}
\begin{IEEEproof} See Appendix \ref{prooflema=8}.
\end{IEEEproof}

Then the only possible class of $(8,8)$ absorption set would have connectivity $[5,5,5,5,5,5,5,5]$.
We claim that
\begin{claim}\label{claima=8}
Graphically, there exist five possible topologies
for $[5,5,5,5,5,5,5,5]$ absorption sets, shown in \figurename \ref{figa=8b488:sub:b}, \ref{figa=8b488:sub:g} and \ref{fig12:sub}.
\end{claim}
\begin{IEEEproof} See Appendix \ref{proofclaima=8}.
\end{IEEEproof}

\begin{thm}
The number of $(8,8)$ absorption sets is $14,272$ and they all have the topology of \figurename \ref{fig12:sub:b}.
\end{thm}
\begin{IEEEproof} By searching all topologies in Claim \ref{claima=8} on the $\mathbf{H}$ matrix of \cite{DjuXuAbdLin03}.
\end{IEEEproof}

\begin{figure}[!t]
\centering \subfigure[] 
{
    \label{fig13:sub:a}
$\xymatrix@M=0pt@W=0pt@R=20pt@C=-2pt
{
&&&&& \color{blue}\boxplus\ar@{--}[d]&&&& \color{blue}\boxplus\ar@{--}[d]&&&& \color{blue}\boxplus\ar@{--}[d]&&&& \color{blue}\boxplus\ar@{--}[d]&&&& \color{blue}\boxplus\ar@{--}[d]&&&& \color{blue}\boxplus\ar@{--}[d]&&&& \color{blue}\boxplus\ar@{--}[d]&&&& \color{blue}\boxplus\ar@{--}[d]&&&&& \\
&&&&& \CIRCLE\ar@{-}[ddlllll]\ar@{-}[ddlll]\ar@{-}[ddl]\ar@{-}[ddr]\ar@{-}[ddrrr]&&&& \CIRCLE\ar@{-}[ddr]\ar@{-}[ddrrr]\ar@{-}[ddrrrrr]\ar@{-}[ddrrrrrrr]\ar@{-}[ddrrrrrrrrrrrrrrrrrrrrr]&&&& \CIRCLE\ar@{-}[ddlllllllllllll]\ar@{-}[ddlll]\ar@{-}[ddrrrrr]\ar@{-}[ddrrrrrrrrrrrrrrrrrrr]\ar@{-}[ddrrrrrrrrr]&&&& \CIRCLE\ar@{-}[ddlllllllllllllll]\ar@{-}[ddlllll]\ar@{-}[ddrrr]\ar@{-}[ddrrrrrrrrrrrrrrrrr]\ar@{-}[ddrrrrrrr]&&&& \CIRCLE\ar@{-}[ddlll]\ar@{-}[ddlllllll]\ar@{-}[ddlllllllllllllllll]\ar@{-}[ddrrrrrrrrrrrrrrr]\ar@{-}[ddrrrrr]&&&& \CIRCLE\ar@{-}[ddrrrrrrrrrrrrr]\ar@{-}[ddrrr]\ar@{-}[ddlllll]\ar@{-}[ddlllllllllllllllllll]\ar@{-}[ddlllllllll]&&&& \CIRCLE\ar@{-}[ddl]\ar@{-}[ddlll]\ar@{-}[ddlllll]\ar@{-}[ddlllllll]\ar@{-}[ddlllllllllllllllllllll]&&&& \CIRCLE\ar@{-}[ddlll]\ar@{-}[ddl]\ar@{-}[ddr]\ar@{-}[ddrrr]\ar@{-}[ddrrrrr]&&&&& \\
&&&&&&&&&&&&&&&&&&&&&&&&&&&&&&&&&&&&&&\\
\boxplus& \hspace{9pt}& \boxplus&\hspace{9pt}&
\boxplus&\hspace{9pt}& \boxplus&\hspace{9pt}& \boxplus&\hspace{9pt}&
\boxplus&\hspace{9pt}& \boxplus&\hspace{9pt}& \boxplus&\hspace{9pt}&
\boxplus&\hspace{9pt}& \boxplus&\hspace{9pt}& \boxplus&\hspace{9pt}&
\boxplus&\hspace{9pt}& \boxplus&\hspace{9pt}& \boxplus&\hspace{9pt}&
\boxplus&\hspace{9pt}& \boxplus&\hspace{9pt}& \boxplus&\hspace{9pt}&
\boxplus&\hspace{9pt}& \boxplus&\hspace{9pt}& \boxplus }$ }\\
\subfigure[] 
{
    \label{fig13:sub:b}
$\xymatrix@M=0pt@W=0pt@R=15.78pt@C=15.78pt
{
& \CIRCLE\ar@{-}[r]\ar@{-}[dl]\ar@{-}[ddl]\ar@{-}[drr]\ar@{-}[ddrr]& \CIRCLE \ar@{-}[dll]\ar@{-}[ddll]\ar@{-}[dr]\ar@{-}[ddr]&\\
\CIRCLE \ar@{-}[d]\ar@{-}[ddr]\ar@{-}[ddrr]&&& \CIRCLE\ar@{-}[d]\ar@{-}[ddl]\ar@{-}[ddll]\\
\CIRCLE\ar@{-}[dr]\ar@{-}[drr]&&& \CIRCLE\ar@{-}[dl]\ar@{-}[dll]\\
& \CIRCLE\ar@{-}[r]& \CIRCLE&\\
}$ } \caption{$(8,8)$ absorptions sets.}
\label{fig13:sub} 
\end{figure}
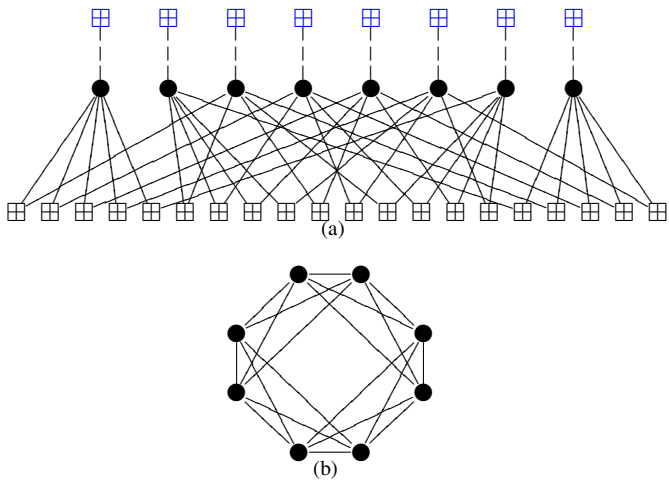

Since these are the dominant absorption sets, let us sketch their
connections in \figurename \ref{fig13:sub} one more time\footnote{It
took approximately ninety minutes on an AMD Opteron Processor
($64$bits/$2.4$GHz) to search the topology \figurename
\ref{fig13:sub}.}.

The average multiplicity of each variable node appeared in such sets is $14272\times 8/2048=55.75$. Because of the block structure of the $\mathbf{H}$ matrix, certain groups of variable nodes do share the same multiplicity, as listed in Table \ref{table2}. The ratio $b/a=1$. The next possible absorption set with these parameters would be the $(10,10)$ set.

\begin{table}[!t]
\renewcommand{\arraystretch}{1.1}
\caption{The multiplicity of each variable node in $(8,8)$ absorption sets.}
\label{table2}
\centering
\begin{tabular}{|c|c||c|c|}\hline
\bfseries Variable Nodes & \bfseries Multiplicities & \bfseries Variable Nodes &
\bfseries Multiplicities\\\hline\hline
$0$---$63$&$63$&$1024$---$1087$&$54$\\\hline
$64$---$127$&$55$&$1088$---$1151$&$66$\\\hline
$128$---$191$&$74$&$1152$---$1215$&$49$\\\hline
$192$---$255$&$60$&$1216$---$1279$&$40$\\\hline
$256$---$319$&$67$&$1280$---$1343$&$75$\\\hline
$320$---$383$&$68$&$1344$---$1407$&$41$\\\hline
$384$---$447$&$54$&$1408$---$1471$&$56$\\\hline
$448$---$511$&$60$&$1472$---$1535$&$59$\\\hline
$512$---$575$&$57$&$1536$---$1599$&$47$\\\hline
$576$---$639$&$66$&$1600$---$1663$&$36$\\\hline
$640$---$703$&$60$&$1664$---$1727$&$59$\\\hline
$704$---$767$&$49$&$1728$---$1791$&$59$\\\hline
$768$---$831$&$39$&$1792$---$1855$&$37$\\\hline
$832$---$895$&$47$&$1856$---$1919$&$52$\\\hline
$896$---$959$&$61$&$1920$---$1983$&$69$\\\hline
$960$---$1023$&$62$&$1984$---$2047$&$43$\\\hline
\end{tabular}
\end{table}

\subsection{Less Dominant Absorption Sets}

There exist larger and less dominant absorption sets. See Appendix \ref{less} for details.

\section{Dynamic Analysis of Absorption Sets}

We now present a linearized analysis to gain insight into the behavior of dominant absorption sets starting with the leading $(8,8)$ absorption set.
First we note that the variable nodes perform simple
addition. Furthermore, the check nodes basically choose the minimum
of the incoming signals. If we make the reasonable assumption that the absorption set converges slower than the remaining nodes in the code, and due to the
fact that each (satisfied) check node is connected exactly to two absorption set variables, the minimum
absolute-value signal into the participating check nodes will come from one of the absorption set variables.
If this is true, the check nodes simple exchange the signals on the connections to the absorption set variable nodes. We will refine this approximation below.

Additionally, each absorption set variable node is singly connected to a ``lone'' floating extrinsic parity
check node, all of whose other connections go to other, set-external variable nodes.
The messages through these eight extrinsic check nodes are the extrinsic messages into the absorption
set, and are of crucial importance. Algorithmically, they play exactly the same role as the intrinsic
channel values which are fed into the variable nodes by virtue of the summation function executed at
the variable nodes.

\begin{figure}[!t]
\centering
\setlength{\unitlength}{1mm}
\begin{picture}(130,53)
\put(0,0){\includegraphics[scale=0.414]{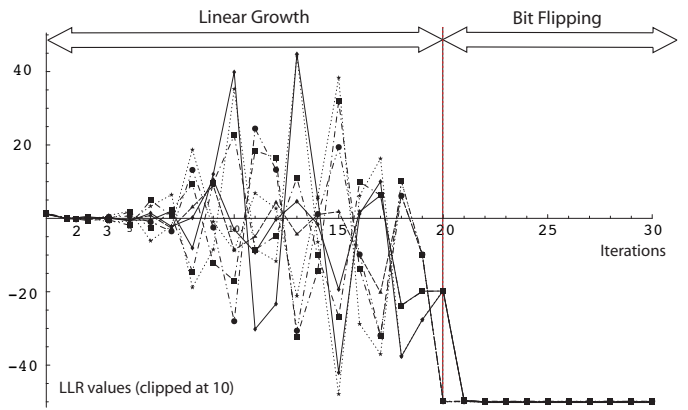}}
\end{picture}
\caption{Dynamics of an absorption set close to the decision boundary. The different curves are the variable node LLR values for the eight absorption set nodes.}
\label{fig:dynamics}
\end{figure}

\figurename \ref{fig:dynamics} shows an example of the dynamic behavior of the absorption set variables
close to its decision threshold boundary. The seemingly erratic behavior resolves after a number
of iterations when all variables follow highly correlated trajectories. This observation is the basis
for the following analysis.

\newcounter{MYtempeqncnt}

\begin{figure*}[!b]
\normalsize \setcounter{MYtempeqncnt}{\value{equation}}
\setcounter{equation}{6} \vspace*{4pt} \hrulefill
\begin{equation}\label{eq:absorptionset}
P_{\rm AS} = Q \left(\frac{\displaystyle 2 m_\lambda
        + 2 \sum\limits_{j=1}^I \left(\frac{m_{\lambda^{(ex)}}^{(j)}+ m_\lambda}{\mu_{\rm max}^j}
    \prod\limits_{l=1}^j \frac{1}{g_l}\right)}
    {\sqrt{  \displaystyle \left( 1 + \sum_{j=1}^I  \prod_{l=1}^j \frac{1}{g_l \mu_{\rm max} }  \right)^2
    m_\lambda + \sum_{j=1}^I  m_{\lambda^{(ex)}}^{(j)}
            \left( \prod_{l=1}^j \frac{1}{g_l \mu_{\rm max}} \right)^2 } } \right)
\end{equation}
\setcounter{equation}{\value{MYtempeqncnt}}
\end{figure*}

Denote the {\em outgoing solid edge values} from the variable nodes (\figurename \ref{fig:trapping}) by $x_i$, i.e., $x_1,\cdots, x_5$ leave variable node $v=0$, $x_6,\cdots, x_{10}$ variable
node $v=1$, etc. Collect the $x_i$ in the length-$40$ column vector $\mathbf{x}$, which is the vector of
outgoing variable edge values in the absorption set. Likewise, and analogously, let $\mathbf y$ be
the {\em incoming edge values} to the variable nodes, such that $y_j$ corresponds to the reverse-direction message. Now, at iteration $i=0$
\begin{displaymath}
\mathbf{x}_0 = \boldsymbol{\lambda}
\end{displaymath}
where the initial input is the vector of channel intrinsics $ \boldsymbol{\lambda} =
\left[\lambda_1, \cdots, \lambda_1, \lambda_2, \cdots, \lambda_2, \cdots, \lambda_8, \cdots, \lambda_8\right]^\mathrm{T}$
duplicated onto the outgoing messages. It undergoes the
following operation at the check node:
\begin{displaymath}
\mathbf{y}_0 = \mathbf{C x}
\end{displaymath}
where $ \mathbf{C}$ is a permutation matrix that exchanges the absorption set signals as discussed
above. At iteration $i=1$ we obtain
\begin{displaymath}
\mathbf{x}_1 = \mathbf{V C}\boldsymbol{\lambda} + \boldsymbol{\lambda}
        + \boldsymbol{\lambda}_1^{(ex)}
\end{displaymath}
where  $\mathbf{V}$ is the variable node function matrix, i.e., each output is the sum
of the other four inputs from the check nodes plus the intrinsic input. The extrinsic inputs
from the remainder of the code graph are contained in  $\boldsymbol{\lambda}_1^{(ex)}$.
Following the linear model, at iteration $I=j$ extrinsic signals are injected into the
absorption set as
$\boldsymbol{\lambda}^{(ex)}_j= \left[ \lambda^{(ex)}_{j1}, \cdots, \lambda^{(ex)}_{j1},
 \lambda^{(ex)}_{j2}, \cdots,  \lambda^{(ex)}_{j8}\right]^\mathrm{T}$ via the extrinsic check nodes.
By induction we obtain at iteration $i=I$
\begin{displaymath}
\mathbf{x}_I = \sum_{i=0}^I (\mathbf{V C})^i  \boldsymbol{\lambda}
+\sum_{j=1}^I \sum_{i=j}^I (\mathbf{V C})^i  \boldsymbol{\lambda}_j^{(ex)}
\end{displaymath}
Applying the spectral theorem we obtain
\begin{displaymath}
 (\mathbf{V C})^i  \mathbf{\lambda}  \rightarrow \mu_{\rm max}^i
 \left( \mathbf{\lambda}^\mathrm{T}  \mathbf{v}_{\rm max} \right) \mathbf{v}_{\rm max}
\end{displaymath}
where $\mathbf{v}_{\rm max}$ is the unit-length eigenvector of the maximal eigenvalue
$\mu_{\rm max}$ of the matrix $\mathbf{V C}$.

The following lemma holds:
\begin{lem}
The largest eigenvalue of $\mathbf{V C}$ for the $(8,8)$ set is $\mu_{\rm max} = d_v-2=4$, and its associated eigenvector
is $\mathbf{v}_{\rm max} = [1,\cdots, 1]^\mathrm{T}$.
\end{lem}

\begin{IEEEproof} First write $\mathbf{V C} = 4 \mathbf{M}$. By inspection $\mathbf{M}$ is a probability
matrix, i.e., the sum of all rows equals unity. As a special case of the Perron-Frobenius theorem it
is known that the largest eigenvalue of a probability matrix is $1$, therefore the largest eigenvalue
of $\mathbf{V C}$ equals $4$.
By inspection $\mathbf{V C} [1,\cdots,1]^\mathrm{T} = 4 [1,\cdots,1]^\mathrm{T}$.
\end{IEEEproof}

The absorption set in question falls in error if
\begin{equation}\label{eq1}
\beta=\boldsymbol{\lambda}^\mathrm{T}  \mathbf{v}_{\rm max}
+ \sum_{j=1}^I \frac{ \left( {\boldsymbol{\lambda}_j^{(ex)}}  + \boldsymbol{\lambda} \right)^\mathrm{T}
\mathbf{v}_{\rm max} }
    {\mu_{\rm max}^j} \leq 0
\end{equation}
or, in the case of the $(8,8)$ absorption set
\begin{equation}
\beta=\sum_{i=1}^8 \left( \lambda_i + \sum_{j=1}^I \frac{ \lambda^{(ex)}_{ji} + \lambda_i }
    {\mu_{\rm max}^j}   \right) \leq 0
\label{eq:TSconvergence}
\end{equation}
The eigenvalue $\mu_{\rm max}=d_v-2$ is the {\em gain} of the absorption
set and it is determined by the variable node degree.

Exact knowledge of $\lambda^{(ex)}_{ji}$ is not available to the analysis, since these values depend
on the received signals. However, assuming that the code structure extrinsic to the apsorption set
operates ``regularly'', we may substitute average values for the $\lambda^{(ex)}_{ji}$. Note that
$\lambda_i$ is Gaussian distributed from the channel, and that we may assume that $\lambda^{(ex)}_{ji}$ is also Gaussian distributed as is customary in density evolution analysis \cite{ChuRicUrb01,SchPer04}. Furthermore, like $\lambda_i$, we assume
that $\lambda^{(ex)}_{ji}$ has a {\em consistent} Gaussian distribution with $m=2 \sigma^2$, where
$m$ is the mean. We therefore only need the mean of $\lambda^{(ex)}_{ji}$, which we can
calculate from a Gaussian density evolution calculation\footnote{For details and definitions, see \cite[Chapter 11]{SchPer04}.}, i.e.,
\begin{displaymath}
m_{\lambda^{(ex)}}^{(i)} = \phi^{-1} \left(1 - \left[1-
    \phi\left( m_\lambda + (d_v-1) m_{\lambda^{(ex)}}^{(i-1)} \right) \right]^{d_c-1} \right)
\end{displaymath}
where $m_\lambda = 2 E_b/\sigma^2$ is the mean of $\lambda_i$, $m_{\lambda^{(ex)}}^{(i)}$ is the mean of the extrinsic signal $\lambda^{(ex)}_{ji}$, and $\phi$ is the check node mean transfer function \cite{SchPer04}.

With the Gaussian assumptions, the probability of (\ref{eq:TSconvergence}) happening
can be calculated as
\setlength{\arraycolsep}{0.0em}
\begin{eqnarray}
P_{\rm AS} &{ }={ }&
{\rm Pr} \left( \beta \leq 0 \right)  \nonumber\\[1mm]
& = & Q \left( \frac{\displaystyle 2 m_\lambda
        + 2 \sum\limits_{j=1}^I \frac{m_{\lambda^{(ex)}}^{(j)}+ m_\lambda}{\mu_{\rm max}^j} }
    {\sqrt{ \displaystyle \left( 1 +  \sum_{j=1}^I  \frac{1}{\mu_{\rm max}^j }  \right)^2
    m_\lambda + \sum_{j=1}^I  \frac{m_{\lambda^{(ex)}}^{(j)}}{\mu_{\rm max}^{2j}} } } \right)
\label{eq:TSerror}
\end{eqnarray}
\setlength{\arraycolsep}{5pt}

Two refinements can be added to this analysis. The exchange of
extrinsics through the matrix $\mathbf{C}$ is an approximation in
two ways: (i) As long as the remaining $d_c-2=30$ inputs to the
check node are relatively small, the entries of $\mathbf{C}$ are
strictly less than unity, and, (ii) in case one of the extrinsic
incoming check node messages has the wrong polarity, the returned
signal to the absorption set switches polarity. Case (i) is
approached as follows. Using a Taylor series approximation we show
that the quintessential check node operation
\begin{equation*}
\tanh^{-1} \left( \tanh(x) \prod_{i=1}^{d_c-2} \tanh(x_i) \right) = \prod_{i=1}^{d_c-2} \tanh(x_i)\, x + O\left[x^3\right]
\end{equation*}
where $\prod\limits_{i=1}^{d_c-2} \tanh(x_i)$ can be interpreted as a ``check node gain''.
If we use for $x_i$ the mean $m_{\mu^{(ex)}}^{(i)}$ of the signals $\mu^{(ex)}$ from the variable
to the check nodes, an average gain can be computed as
\setlength{\arraycolsep}{0.0em}
\begin{eqnarray}
g_i &{ }={ }& E \left[
\prod_{i=1}^{d_c-2} \tanh\left(\frac{m_{\mu^{(ex)}}^{(i)}}{2}\right)\right] \\[1mm]
& = &  E \left[\tanh\left(\frac{m_{\mu^{(ex)}}^{(i)}}{2}\right)\right]^{d_c-2} \\ [3mm]
& = & \left( 1 - \phi\left(m_{\mu^{(ex)}}^{(i)}\right) \right)^{d_c-2}
\end{eqnarray}
\setlength{\arraycolsep}{5pt}
\hspace{-3.4pt}where the last equality results from the definition of the density evolution function
$\phi(\cdot)$.
With this result the probability in (\ref{eq:TSerror}) is modified to
(\ref{eq:absorptionset}). In the case of
general sets we need to work with (\ref{eq1}) instead, and compute $\mu_{\rm max}$ and $\mathbf{v}_{\rm max}$
numerically using the set topology.


\addtocounter{equation}{1}


\begin{figure*}[!b]
\normalsize
\setcounter{MYtempeqncnt}{\value{equation}}
\setcounter{equation}{10}
\vspace*{4pt}
\hrulefill
\begin{equation}\label{eq:f(a,b)}
f(a,b) =
    \frac{\displaystyle a m_\lambda \left(1+ \sum_{j=1}^I \prod_{l=1}^j\frac{1}{g_l \left(5-\frac{b}{a}\right)^j } \right)
        + b \sum\limits_{j=1}^I \left(\frac{m_{\lambda^{(ex)}}^{(j)}}{\left(5-\frac{b}{a}\right)^j}\prod\limits_{l=1}^j \frac{1}{g_l}\right)}
    {\displaystyle  \sqrt{  \displaystyle 2a m_\lambda \left( 1 + \sum_{j=1}^I  \prod_{l=1}^j \frac{1}{g_l \left(5-\frac{b}{a}\right) }  \right)^2
     + 2b\sum_{j=1}^I  m_{\lambda^{(ex)}}^{(j)}
            \left( \prod_{l=1}^j \frac{1}{g_l \left(5-\frac{b}{a}\right)} \right)^2 } }
\end{equation}
\setcounter{equation}{\value{MYtempeqncnt}}
\end{figure*}

Case (ii) can be handled by the linear analysis as well in the following way. If an external variable
to the absorption set has an incorrect sign, this reverses the polarity of the signal returned to
the absorption set from that particular check node. During the first iteration, these extrinsic signals
are basically the received channel LLRs from the connected variable nodes. The probability that
these are in error is given by the raw bit error rate
\begin{equation}
P_e = Q\left( \sqrt{ 2 \frac{E_s}{N_0} } \right)
\end{equation}
There are $d_c-2=30$ external inputs impinging on each check node of the absorption set, therefore
the probability that a returned signal experiences a polarity reversal is given by
\begin{equation}
P_p =  \sum_{k=1}^{14}{30\choose 2k+1}  P_e^{2k+1}
    \left( 1 - P_e \right)^{29-2k}
\end{equation}
The model in (\ref{eq:absorptionset}) can now be expanded by injecting a correction value
into the absorption set node whenever an external value is in error. We assume that if a polarity
reversal occurs, the minimum value of the check node is likely close to zero, therefore the
injected correction value needs
to cancel the absent feedback signal and is set to
$- \lambda_{{\rm ex},i}/\mu_{\rm max}$.
If $k$ check nodes are in error, $2k$ correction values are injected, one for each message
going back to the absorption set. The injected correction values will alter the mean value of
the decision variable to
\begin{equation*}
{\rm mean} \rightarrow 8 \left( m_\lambda \left(1- \frac{k}{4 \mu_{\rm max}} \right)
    +  \sum_{j=1}^I \left(\frac{m_{\lambda^{(ex)}}^{(j)}}{\mu_{\rm max}^j}  \prod_{l=1}^j \frac{1}{g_l}\right) \right)
\end{equation*}
and the variance is adjusted accordingly, where care needs to be taken how the correction values
accumulate. We have used an upper bound on the variance.

Note that these modifications only include check node polarity reversal at the first iteration,
but an extension to subsequent iterations is straight-forward if messy. Furthermore, as seen in
\figurename \ref{fig:trappingsetsim2} (dashed curves), the addition of this mechanism has only a minor
effect on the results.

The probability $P_{\rm AS}$ needs to be multiplied with the multiplicity factor of $14,272$ in order
to obtain a union bound. In order to compute a BER estimate, we further
multiply this number by $8/1723$, since there are eight errors that occur in a frame of $1,723$ bit errors
due to this absorption set.

\figurename \ref{fig:f[a,b]} shows $P_{\rm AS}$ for the first most dominant absorption sets. Also shown
are general tendencies of $P_{\rm AS}$ as a function of $a$ and $b$:
\begin{equation}
P_{\rm AS}=Q\left(f(a,b)\right)
\end{equation}
where $f(a,b)$ is defined by (\ref{eq:f(a,b)}). It can be shown that
$\mu_{\rm max}\approx5-b/a$ is a close approximation (exact for
symmetric sets with $a=b$) to the gain of the set, and this was used
in \figurename \ref{fig:f[a,b]} to plot the curves.

It can be seen that the $(8,8)$ absorption set is the most dominant, which is consistent with numerical observations.
Multiplicities also affect a set's impact --- see \figurename \ref{fig:trappingsetsim2}. Additionally, some sets, like the majority
of $(7,12)$ sets, are ``contained'' in larger sets, that is, such $(7,12)$ absorption sets are not stable under bit flipping and will evolve
into $(8,8)$ sets, of which they are subgraphs.

\begin{figure}[!t]
\centering
\includegraphics[scale=0.41]{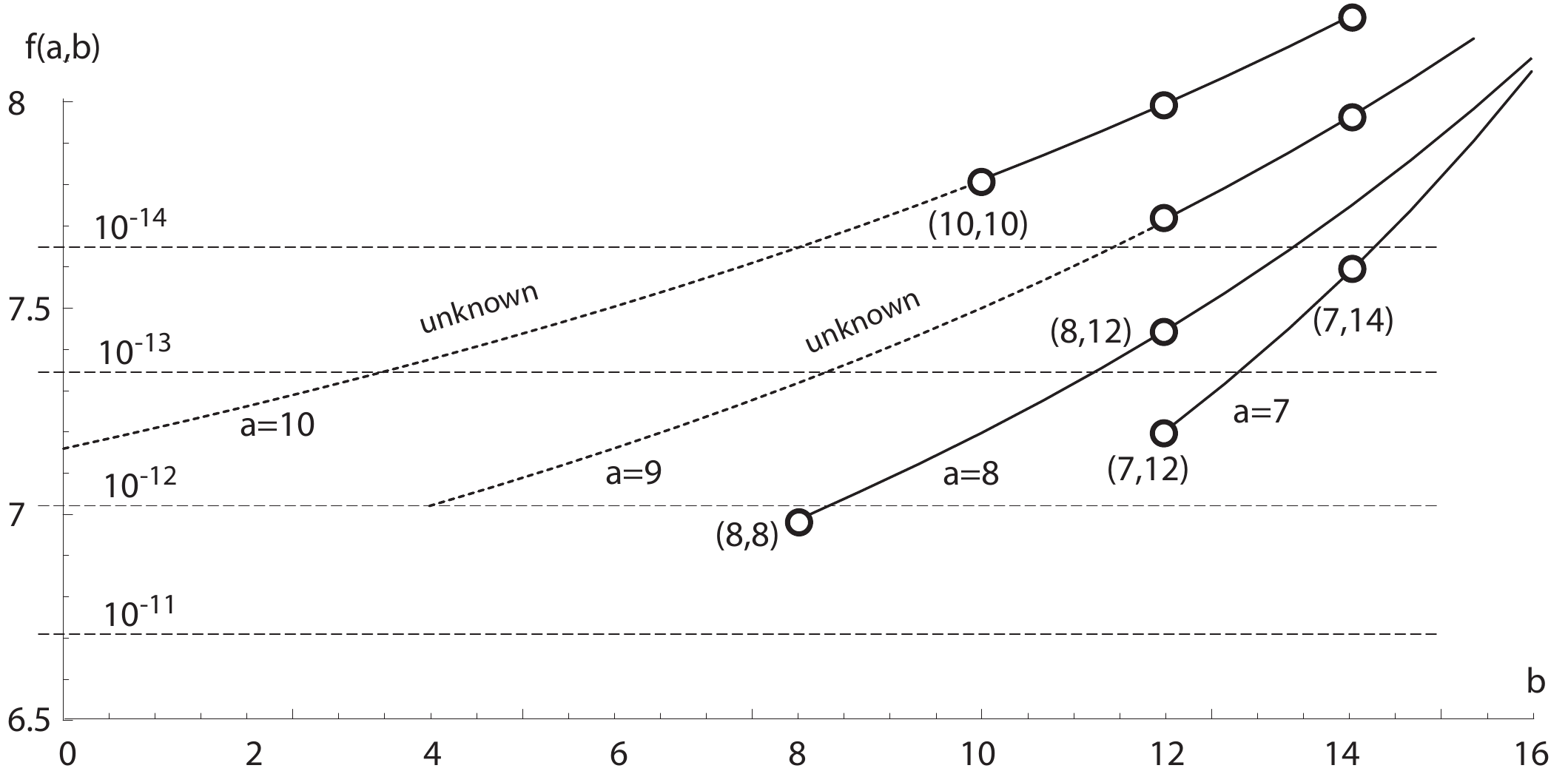}
\caption{Error probability of dominant absorption sets at
$E_b/N_0=5$dB and approximation functions based on $a$ and $b$.
(Curves are drawn only for possible or
 existing parameter combinations.)}
\label{fig:f[a,b]}
\end{figure}

\section{Numerical Verification}

\begin{figure}[!t]
\begin{center}
\setlength{\unitlength}{0.9mm}
\begin{picture}(170,68)
\put(1,0){\includegraphics[scale=0.537]{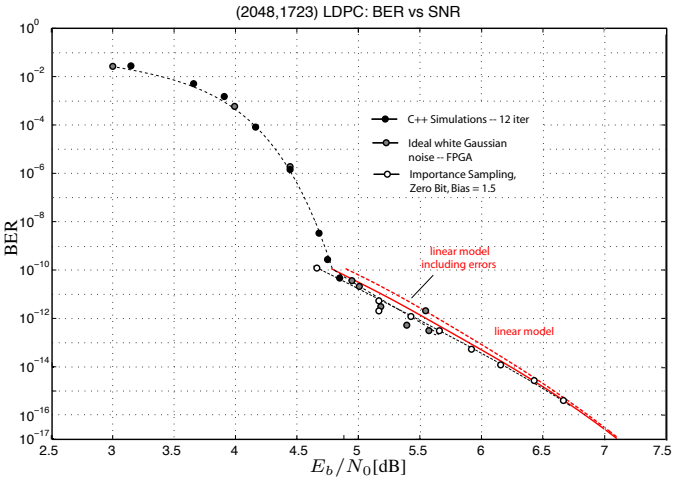}}
\put(45,-2){\scriptsize ${E_b}/{N_0}$[dB]}
\put(0,30){\rotatebox{90}{\scriptsize BER}}
\end{picture}
\end{center}
\caption{IS simulations, FPGA hardware simulations, and analytical error floor analysis for
the $(2048,1723)$ regular $(6,32)$ LDPC code.}
\label{fig:trappingsetsim2}
\end{figure}

\figurename \ref{fig:trappingsetsim2} shows the analytical error
floor calculation using (\ref{eq:absorptionset}) and the
multiplicity of $14,272$. Lesser absorption sets have an impact more
than an order of magnitude lower. And they are not considered. The
figure also shows hardware simulations using an FPGA platform, as
well as importance sampled simulations using the same absorption
sets as bias targets. Regular mean-shift importance sampling was
utilized and each of the absorption sets containing a specific
variable node was biased separately. As evidenced by the figure, our
linearized analysis provides an accurate picture of the error floor
behavior of this code and illustrates the dominance of the $(8,8)$
absorption sets.

\section{Conclusion}

We have presented an analytical analysis of the dynamic behavior of
the dominant absorption sets in LDPC message-passing decoders. These
absorption sets cause the infamous error floor at high
signal-to-noise ratios, and we have identified the dominant such
sets for the example regular LDPC code used in the IEEE 802.3an
standard via topological arguments and searches. Using importance
sampling with the dominant sets accurately predicts the error floor
of this code.


%

\appendices
\section{Proof of Lemma \ref{lema=5}}\label{proofa=5}

Matrix $\mathbf{H}$ is searched observing the constraints imposed by the absorption set topology. In
addition, some of the properties listed in \cite{Zhanetal07, Zhanetal2008} for
array-based LDPC codes apply, as well. The following algorithm
is used:

\begin{algorithm}[H]
\SetVline
\KwIn{Parity-check matrix $\mathbf{H}_{384\times 2048}$ and
\figurename \ref{fig1:sub}.} \KwOut{Absorption sets.}

\ForEach{ variable node $v\in\{0,1,2,\ldots,2047\}$ }{ Pick $4$ out
of $6$ neighboring check nodes  of $v$ denoted $c_0,c_1,c_2$ and $c_3$,
respectively\; \ForEach{ one out of $31$ neighboring variable nodes
other than $v$ of $c_0$, denoted as $v_1$ }{ \ForEach{one out of
$31$ neighboring variable nodes other than $v$ of $c_1$, denoted as
$v_2$}{ \eIf{ $v_2$ and $v_1$ are not connected }{ re-pick $v_2$\;
}{ \ForEach{ one out of $31$ neighboring variable nodes other than
$v$ of $c_2$, denoted as $v_3$ }{\eIf{ $v_3$ and $v_1$ or $v_3$ and
$v_2$ are not connected }{ re-pick $v_3$\; }{ \ForEach{ one out of
$31$ neighboring variable nodes other than $v$ of $c_3$, denoted as
$v_4$ }{\eIf{ $v_4$ and $v_1$ or $v_4$ and $v_2$ or $v_4$ and $v_3$
are not connected }{ re-pick $v_4$\; }{ follow Definition \ref{def1}
to determine if this candidate set is an absorption set\; } } } } }
} } } \caption{Algorithm for finding $(5,10)$ absorption sets.}
\label{alg:mine}
\end{algorithm}

\section{Proof of Theorem \ref{thma=6}}\label{proofa=6}

\begin{figure}[!t]
\centering \subfigure[$(6,6)$] 
{
    \label{fig2:sub:a}
$\xymatrix@M=0pt@W=0pt@R=25pt@C=20pt
{
& \CIRCLE\ar@{-}[r]\ar@{-}[dl]\ar@{-}[dd]\ar@{-}[ddr]\ar@{-}[drr]& \CIRCLE \ar@{-}[dll]\ar@{-}[ddl]\ar@{-}[dd]\ar@{-}[dr]&\\
\CIRCLE \ar@{-}[dr]\ar@{-}[drr]\ar@{-}[rrr]&&& \CIRCLE\ar@{-}[dll]\ar@{-}[dl]\\
& \CIRCLE\ar@{-}[r]& \CIRCLE&\\
}$ }\qquad
\subfigure[$(6,8)$] 
{
    \label{fig2:sub:b}
$\xymatrix@M=0pt@W=0pt@R=25pt@C=20pt
{
& \CIRCLE\ar@{-}[dl]\ar@{-}[dd]\ar@{-}[ddr]\ar@{-}[drr]& \CIRCLE \ar@{-}[dll]\ar@{-}[ddl]\ar@{-}[dd]\ar@{-}[dr]&\\
\CIRCLE \ar@{-}[dr]\ar@{-}[drr]\ar@{-}[rrr]&&& \CIRCLE\ar@{-}[dll]\ar@{-}[dl]\\
& \CIRCLE\ar@{-}[r]& \CIRCLE&\\
}$ }\\
\subfigure[$(6,10)$] 
{
    \label{fig2:sub:c}
$\xymatrix@M=0pt@W=0pt@R=25pt@C=20pt
{
& \CIRCLE\ar@{-}[dl]\ar@{-}[dd]\ar@{-}[ddr]\ar@{-}[drr]& \CIRCLE \ar@{-}[dll]\ar@{-}[ddl]\ar@{-}[dd]\ar@{-}[dr]&\\
\CIRCLE \ar@{-}[dr]\ar@{-}[drr]&&& \CIRCLE\ar@{-}[dll]\ar@{-}[dl]\\
& \CIRCLE\ar@{-}[r]& \CIRCLE&\\
}$ }\qquad
\subfigure[$(6,12)$] 
{
    \label{fig2:sub:d}
$\xymatrix@M=0pt@W=0pt@R=25pt@C=20pt
{
& \CIRCLE\ar@{-}[dl]\ar@{-}[dd]\ar@{-}[ddr]\ar@{-}[drr]& \CIRCLE \ar@{-}[dll]\ar@{-}[ddl]\ar@{-}[dd]\ar@{-}[dr]&\\
\CIRCLE \ar@{-}[dr]\ar@{-}[drr]&&& \CIRCLE\ar@{-}[dll]\ar@{-}[dl]\\
& \CIRCLE& \CIRCLE&\\
}$ } \caption{Possible topologies of size-$6$ absorption sets.}
\label{fig2:sub} 
\end{figure}
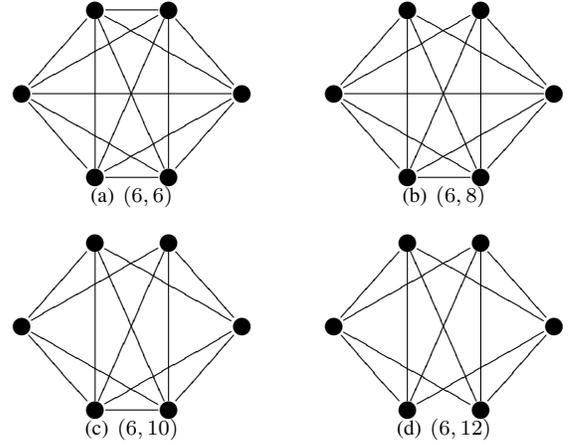

For $a=6$, there are four possible values for $b$ and there is only
one possible topology corresponding to each of them, shown in
\figurename \ref{fig2:sub}.

By removing any node in \figurename \ref{fig2:sub:a} or either
degree-$4$ node in \figurename \ref{fig2:sub:b}, we obtain the
$[4,4,4,4,4]$ set, one shown in \figurename
\ref{fig1:sub:b}. By Lemma \ref{lema=5}, \figurename
\ref{fig2:sub:a}--\ref{fig2:sub:b} do not exist. After eliminating topologies
\figurename \ref{fig2:sub:c}--\ref{fig2:sub:d} algorithmically,
Theorem \ref{thma=6} follows.

\section{Proof of Theorem \ref{thma=7}}\label{proofa=7}

Now $a$ is large enough for the neighboring check nodes to be connected to the absorption set four times. First, we suppose that all satisfied check nodes are connected to the set twice.

\subsection{$b=0$}
If $b=0$, then the absorption set is a codeword. However, since
$d_{\rm min}\geq 8$ \cite{DjuXuAbdLin03}, $b\not= 0$.

\subsection{$0<b<12$}

We apply the constraints in Definition \ref{constraints} and the
pigeonhole principle to prove this.
\begin{enumerate}\setlength{\itemsep}{0pt}
\item $b=2$, there are two classes:
  \begin{enumerate}\setlength{\itemsep}{0pt}
    \item $[6,6,6,6,6,5,5]$: removing either degree-$5$ node leaves a $(6,6)$ absorption set.
    \item $[6,6,6,6,6,6,4]$: removing the degree-$4$ node generates a $(6,4)$ absorption set.
  \end{enumerate}
\item $b=4$, there are three classes:
  \begin{enumerate}\setlength{\itemsep}{0pt}
    \item $[6,6,6,5,5,5,5]$: removing any degree-$5$ node generates a
    $(6,8)$ absorption set.
    \item $[6,6,6,6,5,5,4]$: removing the degree-$4$ node generates a
    $(6,6)$ absorption set.
    \item $[6,6,6,6,6,4,4]$: this is infeasible since the group of five degree-$6$ nodes requires ten edges emanating from the group of degree-$4$ nodes.
  \end{enumerate}
\item $b=6$, there are four classes:
  \begin{enumerate}\setlength{\itemsep}{0pt}
    \item $[6,5,5,5,5,5,5]$: removing any degree-$5$ node generates a
    $(6,10)$ absorption set.
    \item $[6,6,5,5,5,5,4]$: removing the degree-$4$ node generates a
    $(6,8)$ absorption set.
    \item $[6,6,6,5,5,4,4]$: each of the degree-$5$ nodes and each of the degree-$6$ nodes needs at least one and two edges emanating from the two degree-$4$ nodes, respectively. That makes eight. So there is no connection between the degree-$4$ nodes. Thus removing either of them generates a $(6,8)$ absorption set.
    \item $[6,6,6,6,4,4,4]$: each of the degree-$6$ nodes needs three edges emanating from the two degree-$4$ nodes. That makes twelve. So there is no connection among the three degree-$4$ nodes. Thus removing any of them generates a $(6,8)$ absorption set.
  \end{enumerate}
\item $b=8$, there are four classes:
  \begin{enumerate}\setlength{\itemsep}{0pt}
    \item $[5,5,5,5,5,5,4]$: removing the degree-$4$ node generates a
    $(6,10)$ absorption set.
    \item $[6,5,5,5,5,4,4]$: let us study the intrinsic
    connections between the two degree-$4$ nodes:

    \begin{figure}[!t]
\centering \subfigure[] 
{
    \label{fig2degree4:sub:a}
$\xymatrix@M=0pt@W=0pt@R=30pt@C=10pt{
&&\CIRCLE\ar@{-}[d]\ar@{-}[dl]\ar@{-}[dll]\ar@{-}[dr]&&&\CIRCLE\ar@{-}[d]\ar@{-}[dr]\ar@{-}[drr]\ar@{-}[dl]&&\\
&&&&&&&\\}$ }\qquad
\subfigure[] 
{
    \label{fig2degree4:sub:b}
$\xymatrix@M=0pt@W=0pt@R=30pt@C=10pt{
&\CIRCLE\ar@{-}[rrr]\ar@{-}[d]\ar@{-}[dl]\ar@{-}[dr]&&&\CIRCLE\ar@{-}[d]\ar@{-}[dr]\ar@{-}[dl]&\\
&&&&&\\}$ }\caption{Possible intrinsic connections between two
degree-$4$ nodes.}
\label{fig2degree4:sub} 
\end{figure}
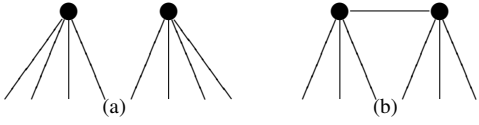

    \begin{enumerate}\setlength{\itemsep}{0pt}
      \item Not connected as shown in \figurename \ref{fig2degree4:sub:a}: removing either degree-$4$ node will reduce it to a $(6,10)$
absorption set.
      \item Connected as shown in \figurename \ref{fig2degree4:sub:b}: we consider the
      connections between the two degree-$4$ nodes and the other five
      nodes in the set. The other five nodes need at least six edges emanating from the two degree-$4$
      nodes, so there is no degree-$5$ node connected to both degree-$4$ nodes. Thus removing both of them generates a $(5,10)$ absorption set.
    \end{enumerate}
    \item $[6,6,5,5,4,4,4]$: the two degree-$5$ nodes and the two degree-$6$ nodes need at least ten edges emanating from the three degree-$4$ nodes. Hence there should be no more than one connection among the group of degree-$4$ nodes:
        \begin{enumerate}\setlength{\itemsep}{0pt}
      \item No connection as shown in \figurename \ref{figinternal3:sub:a}: removing any degree-$4$ node will reduce it to a
$(6,10)$ absorption set.
      \item One connection as shown in \figurename \ref{figinternal3:sub:b}: removing the topmost degree-$4$ node
will reduce it to a $(6,10)$ absorption set.
    \end{enumerate}
    \item $[6,6,6,4,4,4,4]$:  the three degree-$6$ nodes need twelve edges emanating from the four degree-$4$ nodes. Hence there should be no more than two connections among the group of degree-$4$ nodes:
        \begin{enumerate}\setlength{\itemsep}{0pt}
      \item No connection as shown in \figurename \ref{figinternal4:sub:a}: removing any degree-$4$ node will reduce it to a
$(6,10)$ absorption set.
      \item One connection as shown in \figurename \ref{figinternal4:sub:b}: removing either topmost degree-$4$ node will
reduce it to a $(6,10)$ absorption set.
     \item Two connections: there are two cases:
     \begin{enumerate}\setlength{\itemsep}{0pt}
        \item Removing the bottom-right degree-$4$ node in \figurename \ref{figinternal4:sub:c} will
reduce it to a $(6,10)$ absorption set.
        \item Removing either the top or the bottom couple of
degree-$4$ nodes in \figurename \ref{figinternal4:sub:d} will reduce it
to a $(5,10)$ absorption set.
      \end{enumerate}
    \end{enumerate}
  \end{enumerate}
\item $b=10$, there are three classes:
  \begin{enumerate}\setlength{\itemsep}{0pt}
    \item $[5,5,5,5,4,4,4]$: the four degree-$5$ nodes need at least eight edges emanating from the three degree-$4$ nodes. Hence there should be no more than two connections among the group of degree-$4$ nodes:
    \begin{enumerate}\setlength{\itemsep}{0pt}
      \item No connection as shown in \figurename \ref{figinternal3:sub:a}: removing any degree-$4$ node will reduce it to a
$(6,12)$ absorption set.
      \item One connection as shown in \figurename \ref{figinternal3:sub:b}: removing the topmost degree-$4$ node
will reduce it to a $(6,12)$ absorption set.
      \item\label{check1} Two connections as shown in \figurename \ref{figinternal3:sub:c}. No node can be removed to get another absorption set. However, it is straightforward to see that there is only one possible topology to satisfy this:
$$\xymatrix@M=0pt@W=0pt@R=40pt@C=10pt{
& \CIRCLE\ar@{-}[rr]\ar@{-}[drrr]\ar@{-}[dr]\ar@{-}[dl]&& \CIRCLE\ar@{-}[rr]\ar@{-}[drrr]\ar@{-}[dlll]&& \CIRCLE\ar@{-}[dlll]\ar@{-}[dl]\ar@{-}[dr]&\\
\CIRCLE\ar@{-}[rr]\ar@/_0.6pc/@{-}[rrrr]\ar@/_1.2pc/@{-}[rrrrrr]&&
\CIRCLE\ar@{-}[rr]\ar@/_0.6pc/@{-}[rrrr]&& \CIRCLE\ar@{-}[rr]&&
\CIRCLE\\}$$

\bigskip

Therefore we have to go check the $\mathbf{H}$ matrix algorithmically.
    \end{enumerate}
    \item $[6,5,5,4,4,4,4]$: the two degree-$5$ nodes and the degree-$6$ node need at least ten edges emanating from the four degree-$4$ nodes. Hence there should be no more than three connections among the group of degree-$4$ nodes:
    \begin{enumerate}\setlength{\itemsep}{0pt}
      \item No connection as shown in \figurename \ref{figinternal4:sub:a}: removing any degree-$4$ node will reduce it to a
$(6,12)$ absorption set.
      \item One connection as shown in \figurename \ref{figinternal4:sub:b}: removing either topmost degree-$4$ node will
reduce it to a $(6,12)$ absorption set.
     \item Two connections: there are two cases:
     \begin{enumerate}\setlength{\itemsep}{0pt}
        \item Removing the bottom-right degree-$4$ node in \figurename \ref{figinternal4:sub:c} will
reduce it to a $(6,12)$ absorption set.
        \item\label{check2} It is straightforward that there is only one possible topology to satisfy \figurename \ref{figinternal4:sub:d}:
$$\xymatrix@M=0pt@W=0pt@R=40pt@C=10pt{
\CIRCLE\ar@{-}[rr]\ar@{-}[dr]\ar@{-}[drrr]\ar@{-}[drrrrr]&& \CIRCLE\ar@{-}[dl]\ar@{-}[dr]\ar@{-}[drrr]&& \CIRCLE\ar@{-}[rr]\ar@{-}[dr]\ar@{-}[dl]\ar@{-}[dlll]&& \CIRCLE\ar@{-}[dl]\ar@{-}[dlll]\ar@{-}[dlllll]\\
& \CIRCLE&& \CIRCLE\ar@{-}[rr]\ar@{-}[ll]&& \CIRCLE&\\}$$ We have to
turn to $\mathbf{H}$ to show its non-existence.
      \end{enumerate}
      \item Three connections: there are three cases:
      \begin{enumerate}\setlength{\itemsep}{0pt}
\item Removing the bottom-right degree-$4$ node in \figurename \ref{figinternal4:sub:e} will
reduce it to a $(6,12)$ absorption set.
\item\label{check3} It is straightforward that there is only one possible topology to satisfy \figurename \ref{figinternal4:sub:f}:
$$\xymatrix@M=0pt@W=0pt@R=40pt@C=10pt{
\CIRCLE\ar@{-}[rr]\ar@{-}[dr]\ar@{-}[drrr]\ar@{-}[drrrrr]&& \CIRCLE\ar@{-}[dl]\ar@{-}[dr]\ar@{-}[rr]&& \CIRCLE\ar@{-}[rr]\ar@{-}[dr]\ar@{-}[dl]&& \CIRCLE\ar@{-}[dl]\ar@{-}[dlll]\ar@{-}[dlllll]\\
& \CIRCLE\ar@/_0.8pc/@{-}[rrrr]&& \CIRCLE\ar@{-}[rr]\ar@{-}[ll]&&
\CIRCLE&\\}$$

\bigskip We have to turn to $\mathbf{H}$ to show its non-existence.
\item\label{check4} It is straightforward that there is only one possible topology to satisfy \figurename \ref{figinternal4:sub:g}:
$$\xymatrix@M=0pt@W=0pt@R=40pt@C=10pt{
\CIRCLE\ar@{-}[rr]\ar@{-}[dr]\ar@{-}[drrr]\ar@{-}[drrrrr]&& \CIRCLE\ar@{-}[dr]\ar@{-}[rr]\ar@/^0.8pc/@{-}[rrrr]&& \CIRCLE\ar@{-}[dlll]\ar@{-}[dl]\ar@{-}[dr]&& \CIRCLE\ar@{-}[dl]\ar@{-}[dlll]\ar@{-}[dlllll]\\
& \CIRCLE\ar@/_0.8pc/@{-}[rrrr]&& \CIRCLE\ar@{-}[rr]\ar@{-}[ll]&&
\CIRCLE&\\}$$

\medskip We have to turn to $\mathbf{H}$ to show its non-existence.
\end{enumerate}
    \end{enumerate}
    \item\label{check5} $[6,6,4,4,4,4,4]$: it is straightforward to see that there is only one possible topology in this class:
$$\xymatrix@M=0pt@W=0pt@R=40pt@C=30pt{
\CIRCLE\ar@{-}[r]\ar@{-}[dr]\ar@{-}[drrr]\ar@/^0.8pc/@{-}[rrrr]& \CIRCLE\ar@{-}[r]\ar@{-}[d]\ar@{-}[drr]& \CIRCLE\ar@{-}[r]\ar@{-}[dl]\ar@{-}[dr]& \CIRCLE\ar@{-}[r]\ar@{-}[d]\ar@{-}[dll]& \CIRCLE \ar@{-}[dl]\ar@{-}[dlll]\\
& \CIRCLE\ar@{-}[rr]&& \CIRCLE&\\}$$ We have to turn to $\mathbf{H}$
to show its non-existence.
  \end{enumerate}
\end{enumerate}

\begin{figure}[!t]
\centering \subfigure[] 
{
    \label{figinternal3:sub:a}
$\xymatrix@M=0pt@W=0pt@R=20pt@C=7pt{
&\CIRCLE&\\
\CIRCLE&&\CIRCLE\\}$ }\qquad
\subfigure[] 
{
    \label{figinternal3:sub:b}
$\xymatrix@M=0pt@W=0pt@R=20pt@C=7pt{
&\CIRCLE&\\
\CIRCLE\ar@{-}[rr]&&\CIRCLE\\}$ }\qquad
\subfigure[] 
{
    \label{figinternal3:sub:c}
$\xymatrix@M=0pt@W=0pt@R=20pt@C=7pt{
&\CIRCLE\ar@{-}[dl]\ar@{-}[dr]&\\
\CIRCLE&&\CIRCLE\\}$ } \caption{Possible intrinsic connections
among three nodes.}
\label{figinternal3:sub} 
\end{figure}
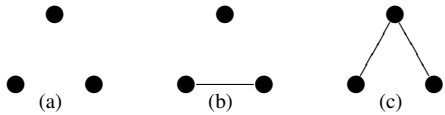

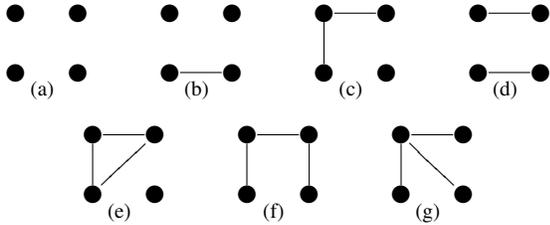
\begin{figure}[!t]
\centering \subfigure[] 
{
    \label{figinternal4:sub:a}
$\xymatrix@M=0pt@W=0pt@R=15.78pt@C=15.78pt{
\CIRCLE&\CIRCLE\\
\CIRCLE&\CIRCLE\\}$ }\qquad
\subfigure[] 
{
    \label{figinternal4:sub:b}
$\xymatrix@M=0pt@W=0pt@R=15.78pt@C=15.78pt{
\CIRCLE&\CIRCLE\\
\CIRCLE\ar@{-}[r]&\CIRCLE\\}$ }\qquad
\subfigure[] 
{
    \label{figinternal4:sub:c}
$\xymatrix@M=0pt@W=0pt@R=15.78pt@C=15.78pt{
\CIRCLE\ar@{-}[r]\ar@{-}[d]&\CIRCLE\\
\CIRCLE&\CIRCLE\\}$ }\qquad
\subfigure[] 
{
    \label{figinternal4:sub:d}
$\xymatrix@M=0pt@W=0pt@R=15.78pt@C=15.78pt{
\CIRCLE\ar@{-}[r]&\CIRCLE\\
\CIRCLE\ar@{-}[r]&\CIRCLE\\}$ }\\
\subfigure[] 
{
    \label{figinternal4:sub:e}
$\xymatrix@M=0pt@W=0pt@R=15.78pt@C=15.78pt{
\CIRCLE\ar@{-}[r]\ar@{-}[d]&\CIRCLE\ar@{-}[dl]\\
\CIRCLE&\CIRCLE\\}$ }\qquad
\subfigure[] 
{
    \label{figinternal4:sub:f}
$\xymatrix@M=0pt@W=0pt@R=15.78pt@C=15.78pt{
\CIRCLE\ar@{-}[r]\ar@{-}[d]&\CIRCLE\ar@{-}[d]\\
\CIRCLE&\CIRCLE\\}$ }\qquad
\subfigure[] 
{
    \label{figinternal4:sub:g}
$\xymatrix@M=0pt@W=0pt@R=15.78pt@C=15.78pt{
\CIRCLE\ar@{-}[r]\ar@{-}[d]\ar@{-}[dr]&\CIRCLE\\
\CIRCLE&\CIRCLE\\}$ } \caption{Possible intrinsic connections among
four nodes.}
\label{figinternal4:sub} 
\end{figure}

\subsection{$b=12$}
The extrinsic degree is large now. We will see in the $a=8$ section
that there are smaller $b$'s and $(7,12)$ absorption sets do exist as a reduction from them.

\subsection{$b=14$}

Table \ref{table714} shows the existence of $(7,14)$ sets.

\begin{table}[!t]
\renewcommand{\arraystretch}{1.1}
\caption{An example of $(7,14)$ absorption set.} \label{table714}
\centering
\begin{tabular}{|c||c|c|c|c|c|c|}\hline
\bfseries Variable Nodes & \multicolumn{6}{c|}{\bfseries Six Neighboring Check Nodes}
\\\hline\hline 0&56&120&184&248&312&376\\\hline
109&56&79&174&199&300&349\\\hline 740&21&120&138&236&310&325\\\hline
1801&21&87&184&202&300&374\\\hline
1765&14&125&169&199&266&376\\\hline 43&10&74&138&202&266&330\\\hline
862&10&125&174&242&285&325\\\hline
\end{tabular}\end{table}

\bigskip

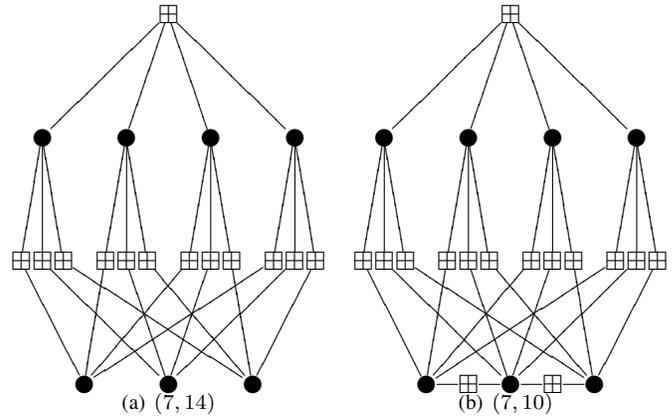
\begin{figure}[!t]
\centering \subfigure[$(7,14)$] 
{
    \label{figb4}
$\xymatrix@M=0pt@W=0pt@R=40pt@C=0.18pt
{
&&&&&&& \boxplus\ar@{-}[dll]\ar@{-}[dllllll]\ar@{-}[drr]\ar@{-}[drrrrrr]&&&&&&&\\
& \CIRCLE\ar@{-}[dl]\ar@{-}[d]\ar@{-}[dr]&&&& \CIRCLE\ar@{-}[dl]\ar@{-}[d]\ar@{-}[dr]&&&& \CIRCLE\ar@{-}[dl]\ar@{-}[d]\ar@{-}[dr]&&&& \CIRCLE\ar@{-}[dl]\ar@{-}[d]\ar@{-}[dr]&\\
\boxplus& \boxplus& \boxplus&& \boxplus& \boxplus& \boxplus&& \boxplus& \boxplus& \boxplus&& \boxplus& \boxplus& \boxplus\\
&&& \CIRCLE\ar@{-}[ulll]\ar@{-}[ur]\ar@{-}[urrrrr]\ar@{-}[urrrrrrrrr]&&&& \CIRCLE\ar@{-}[ull]\ar@{-}[ullllll]\ar@{-}[urr]\ar@{-}[urrrrrr]&&&& \CIRCLE\ar@{-}[ulllllllll]\ar@{-}[ulllll]\ar@{-}[ul]\ar@{-}[urrr]&&&\\
}$ }
\subfigure[$(7,10)$] 
{
    \label{figb41}
$\xymatrix@M=0pt@W=0pt@R=40pt@C=0.18pt
{
&&&&&&& \boxplus\ar@{-}[dll]\ar@{-}[dllllll]\ar@{-}[drr]\ar@{-}[drrrrrr]&&&&&&&\\
& \CIRCLE\ar@{-}[dl]\ar@{-}[d]\ar@{-}[dr]&&&& \CIRCLE\ar@{-}[dl]\ar@{-}[d]\ar@{-}[dr]&&&& \CIRCLE\ar@{-}[dl]\ar@{-}[d]\ar@{-}[dr]&&&& \CIRCLE\ar@{-}[dl]\ar@{-}[d]\ar@{-}[dr]&\\
\boxplus& \boxplus& \boxplus&& \boxplus& \boxplus& \boxplus&& \boxplus& \boxplus& \boxplus&& \boxplus& \boxplus& \boxplus\\
&&& \CIRCLE\ar@{-}[rr]\ar@{-}[ulll]\ar@{-}[ur]\ar@{-}[urrrrr]\ar@{-}[urrrrrrrrr]&& \boxplus&& \CIRCLE\ar@{-}[ll]\ar@{-}[rr]\ar@{-}[ull]\ar@{-}[ullllll]\ar@{-}[urr]\ar@{-}[urrrrrr]&& \boxplus&& \CIRCLE\ar@{-}[ll]\ar@{-}[ulllllllll]\ar@{-}[ulllll]\ar@{-}[ul]\ar@{-}[urrr]&&&\\
}$}
 \caption{One check node connecting to size-$7$ sets four times.}
\label{figbbbb4:sub} 
\end{figure}

If we allow the satisfied check nodes to be connected to the set more more than twice, it is clear that only one check node could be connected to the set four times, as shown in \figurename \ref{figb4}, which is a $(7,14)$ absorption set. Now, there can be no other intrinsic connections among the four variable nodes at the top --- this would create a $4$-cycle. Thus depending on the intrinsic connections among the three variable nodes at the bottom, we could obtain $(7,12)$, $(7,10)$ or $(7,8)$ absorption sets, respectively. Both $(7,14)$ and $(7,12)$ absorption sets exist and the $(7,8)$ can reduce to $(5,10)$ by removing any two degree-$4$ nodes and does not exist therefore. Only $(7,10)$ sets, \figurename \ref{figb41}, need to be searched.

After searching the topologies in \ref{check1}, \ref{check2},
\ref{check3}, \ref{check4}, \ref{check5} and \figurename \ref{figb41} with $\mathbf{H}$ algorithmically, Lemma
\ref{thma=7} follows.

\section{Proof of Lemma \ref{lema=8}}\label{prooflema=8}

Again, we first suppose that all satisfied check nodes are connected to the
set twice. We apply the constraints in Definition \ref{constraints} and the pigeonhole principle
to prove this.

\subsection{$b=0$}
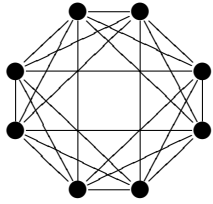
\begin{figure}[!t]
\centering
$\xymatrix@M=0pt@W=0pt@R=15.78pt@C=15.78pt
{
& \CIRCLE\ar@{-}[r]\ar@{-}[dl]\ar@{-}[ddl]\ar@{-}[drr]\ar@{-}[ddrr]\ar@{-}[ddd]& \CIRCLE \ar@{-}[dll]\ar@{-}[ddll]\ar@{-}[dr]\ar@{-}[ddr]\ar@{-}[ddd]&\\
\CIRCLE \ar@{-}[d]\ar@{-}[ddr]\ar@{-}[ddrr]\ar@{-}[rrr]&&& \CIRCLE\ar@{-}[d]\ar@{-}[ddl]\ar@{-}[ddll]\\
\CIRCLE\ar@{-}[dr]\ar@{-}[drr]\ar@{-}[rrr]&&& \CIRCLE\ar@{-}[dl]\ar@{-}[dll]\\
& \CIRCLE\ar@{-}[r]& \CIRCLE&\\
}$ \caption{Only possible topology of $(8,0)$ absorption
sets.}\label{fig7}
\end{figure}

In other words, a class of $[6,6,6,6,6,6,6,6]$ absorption
  sets. We obtain the perfectly symmetric \figurename \ref{fig7} again as \figurename \ref{fig1:sub:b}.
Removing any node will reduce it to a $(7,6)$ absorption set.

\subsection{$b=2$}
There are two classes:
  \begin{enumerate}\setlength{\itemsep}{0pt}
    \item $[6,6,6,6,6,6,5,5]$: removing either degree-$5$ node generates a $(7,6)$ absorption set.
    \item $[6,6,6,6,6,6,6,4]$: removing the degree-$4$ node generates a $(7,4)$ absorption set.
  \end{enumerate}

\subsection{$b=4$}
There are three classes:
  \begin{enumerate}\setlength{\itemsep}{0pt}
    \item $[6,6,6,6,5,5,5,5]$: removing any degree-$5$ node generates a $(7,8)$ absorption set.
    \item $[6,6,6,6,6,5,5,4]$: removing the degree-$4$ node generates a $(7,6)$ absorption set.
    \item $[6,6,6,6,6,6,4,4]$: removing both degree-$4$ nodes generates a $(6,10)$ or a $(6,8)$ absorption set.
  \end{enumerate}

\subsection{$b=6$}
There are four classes:
  \begin{enumerate}\setlength{\itemsep}{0pt}
    \item $[6,6,5,5,5,5,5,5]$: removing any degree-$5$ node generates a $(7,10)$ absorption set.
    \item $[6,6,6,5,5,5,5,4]$: removing the degree-$4$ node generates a $(7,8)$ absorption set.
    \item $[6,6,6,6,5,5,4,4]$: let us study the intrinsic
    connections between the two degree-$4$ nodes:
    \begin{enumerate}\setlength{\itemsep}{0pt}
      \item Not connected as shown in \figurename \ref{fig2degree4:sub:a}: Removing either degree-$4$ node will reduce it to a $(7,8)$
absorption set.
      \item Connected as shown in \figurename \ref{fig2degree4:sub:b}: We consider the
      connections between the two degree-$4$ nodes and the other six
      nodes in the set. There are six edges emanating from the two degree-$4$
      nodes and at least four of the six edges must go to the four
      degree-$6$ nodes, respectively. So at most two edges emanating from
      the two degree-$4$ nodes can be connected to the two degree-$5$
      nodes.
      \begin{enumerate}\setlength{\itemsep}{0pt}
        \item If either of the two degree-$5$ nodes is connected to the two degree-$4$ nodes at most once:
        removing both degree-$4$ nodes generates a $(6,8)$ absorption set.
        \item If one degree-$5$ node is connected to both degree-$4$ nodes: removing the other degree-$5$
        node generates a $(7,10)$ absorption set.
      \end{enumerate}
    \end{enumerate}
    \item $[6,6,6,6,6,4,4,4]$: let us study the intrinsic
    connections among the three degree-$4$ nodes. There are five degree-$6$ nodes, which require ten edges from the
three degree-$4$ nodes. Thus there should be no more than one connection among them.
    \begin{enumerate}\setlength{\itemsep}{0pt}
      \item No connection as shown in \figurename \ref{figinternal3:sub:a}: removing any degree-$4$ node will reduce it to a
$(7,8)$ absorption set.
      \item One connection as shown in \figurename \ref{figinternal3:sub:b}: removing the topmost degree-$4$ node
will reduce it to a $(7,8)$ absorption set.
    \end{enumerate}
  \end{enumerate}

\subsection{$b=8$}
There are four classes by assuming there is a degree-$6$ node:
  \begin{enumerate}\setlength{\itemsep}{0pt}
    \item $[6,5,5,5,5,5,5,4]$: removing the degree-$4$ node
will reduce it to a $(7,10)$ absorption set.
    \item $[6,6,5,5,5,5,4,4]$: let us study the intrinsic
    connections between the two degree-$4$ nodes:
    \begin{enumerate}\setlength{\itemsep}{0pt}
      \item Not connected as shown in \figurename \ref{fig2degree4:sub:a}: Removing either degree-$4$ node will reduce it to a $(7,10)$
absorption set.
      \item Connected as shown in \figurename \ref{fig2degree4:sub:b}: We consider the
      connections between the two degree-$4$ nodes and the other six
      nodes in the set. There are six edges emanating from the two degree-$4$
      nodes and at least two and at most four of the six edges must go to the two
      degree-$6$ nodes, respectively. So at most four and at least two edges emanating from
      the two degree-$4$ nodes can be connected to the four degree-$5$
      nodes.
      \begin{enumerate}\setlength{\itemsep}{0pt}
        \item Two edges between the group of degree-$4$ nodes and the group of degree-$6$ nodes:
$$\xymatrix@M=0pt@W=0pt@R=30pt@C=19pt{
\text{degree-$4$ nodes:}&\CIRCLE\ar@{-}[r]\ar@{-}[d]\ar@{-}[dr]& \CIRCLE&&\CIRCLE\ar@{-}[r]\ar@{-}[d]& \CIRCLE\ar@{-}[d]\\
\text{degree-$6$ nodes:}&\CIRCLE& \CIRCLE&&\CIRCLE& \CIRCLE\\}$$ Note that under the conditions in the above two cases, the two degree-$6$ nodes have to be connected with each other and either of them has to be connected to all the degree-$5$ nodes. In addition, there are four edges coming from the degree-$4$ nodes to the four degree-$5$ nodes. Thus,
\begin{enumerate}\setlength{\itemsep}{0pt}
\item if there is no degree-$5$ node sharing the two degree-$4$ nodes: removing the two degree-$4$ nodes will reduce it to a $(6,10)$ absorption sets.
\item if there is one and at most one degree-$5$ node sharing the two degree-$4$ nodes: the topologies will be fixed, respectively, as
$$\xymatrix@M=0pt@W=0pt@R=40pt@C=25pt{
\color{blue}\CIRCLE\ar@{-}[r]\ar@{-}[d]\ar@{-}[dr]\ar@{-}[drrrrr]& \color{blue}\CIRCLE\ar@{-}[drrr]\ar@{-}[drr]\ar@{-}[drrrr] &&&&\\
\CIRCLE\ar@{-}[r]\ar@/_1pc/@{-}[rr]\ar@/_2pc/@{-}[rrr]\ar@/_3pc/@{-}[rrrr]\ar@/_4pc/@{-}[rrrrr]& \CIRCLE\ar@{-}[r]\ar@/_1pc/@{-}[rr]\ar@/_2pc/@{-}[rrr]\ar@/_3pc/@{-}[rrrr]& \CIRCLE\ar@{-}[r]\ar@/_1pc/@{-}[rr]\ar@/_2pc/@{-}[rrr]& \CIRCLE\ar@{-}[r]& \CIRCLE& \color{blue}\CIRCLE\\
&&&&&\\}$$
$$\xymatrix@M=0pt@W=0pt@R=40pt@C=25pt{
\color{blue}\CIRCLE\ar@{-}[r]\ar@{-}[d]\ar@{-}[drrr]\ar@{-}[drrrrr]& \color{blue}\CIRCLE\ar@{-}[drrr]\ar@{-}[d]\ar@{-}[drrrr] &&&&\\
\CIRCLE\ar@{-}[r]\ar@/_1pc/@{-}[rr]\ar@/_2pc/@{-}[rrr]\ar@/_3pc/@{-}[rrrr]\ar@/_4pc/@{-}[rrrrr]& \CIRCLE\ar@{-}[r]\ar@/_1pc/@{-}[rr]\ar@/_2pc/@{-}[rrr]\ar@/_3pc/@{-}[rrrr]& \CIRCLE\ar@{-}[r]\ar@/_1pc/@{-}[rr]\ar@/_2pc/@{-}[rrr]& \CIRCLE \ar@{-}[r]& \CIRCLE& \color{blue}\CIRCLE\\
&&&&&\\}$$
Removing the two degree-$4$ nodes and that degree-$5$ node will reduce them to $(5,10)$ absorption sets.
\end{enumerate}
        \item Three edges between the group of degree-$4$ nodes and the group of degree-$6$ nodes:
$$\xymatrix@M=0pt@W=0pt@R=30pt@C=19pt{
\text{degree-$4$ nodes:}&\CIRCLE\ar@{-}[r]\ar@{-}[d]\ar@{-}[dr]& \CIRCLE\ar@{-}[dl]\\
\text{degree-$6$ nodes:}&\CIRCLE& \CIRCLE\\}$$ Note that under the conditions in the above case, the two degree-$6$ nodes have to be connected with each other and the bottom-right degree-$6$ node has to be connected to all the degree-$5$ nodes. In addition, there are three edges emanating from the degree-$4$ nodes to the four degree-$5$ nodes. Thus,
  \begin{enumerate}\setlength{\itemsep}{0pt}
\item if there is no degree-$5$ node sharing the two degree-$4$ nodes: removing the two degree-$4$ nodes will reduce it to a $(6,10)$ absorption sets.
\item if there is one and at most one degree-$5$ node sharing the two degree-$4$ nodes: the topology will be fixed as
$$\xymatrix@M=0pt@W=0pt@R=40pt@C=25pt{
\color{blue}\CIRCLE\ar@{-}[r]\ar@{-}[d]\ar@{-}[dr]\ar@{-}[drrrrr]& \color{blue}\CIRCLE\ar@{-}[drrr]\ar@{-}[dl]\ar@{-}[drrrr] &&&&\\
\CIRCLE\ar@{-}[r]\ar@/_1pc/@{-}[rr]\ar@/_2pc/@{-}[rrr]\ar@/_3pc/@{-}[rrrr]& \CIRCLE\ar@{-}[r]\ar@/_1pc/@{-}[rr]\ar@/_2pc/@{-}[rrr]\ar@/_3pc/@{-}[rrrr]& \CIRCLE\ar@{-}[r]\ar@/_1pc/@{-}[rr]\ar@/_2pc/@{-}[rrr]& \CIRCLE\ar@{-}[r]\ar@/_1pc/@{-}[rr]& \CIRCLE& \color{blue}\CIRCLE\\
&&&&&\\}$$
Removing the two degree-$4$ nodes and that degree-$5$ node will reduce it to a $(5,10)$ absorption sets.
\end{enumerate}
        \item Four edges between the group of degree-$4$ nodes and the group of degree-$6$ nodes:
$$\xymatrix@M=0pt@W=0pt@R=30pt@C=19pt{
\text{degree-$4$ nodes:}&\CIRCLE\ar@{-}[r]\ar@{-}[d]\ar@{-}[dr]& \CIRCLE\ar@{-}[d]\ar@{-}[dl]\\
\text{degree-$6$ nodes:}&\CIRCLE& \CIRCLE\\}$$
\begin{enumerate}\setlength{\itemsep}{0pt}
\item if there is no degree-$5$ node sharing the two degree-$4$ nodes: removing the two degree-$4$ nodes will reduce it to a $(6,10)$ absorption sets.
\item if there is one and at most one degree-$5$ node sharing the two degree-$4$ nodes: the topology will be fixed as
$$\xymatrix@M=0pt@W=0pt@R=40pt@C=25pt{
\color{blue}\CIRCLE\ar@{-}[r]\ar@{-}[d]\ar@{-}[dr]\ar@{-}[drrrrr]& \color{blue}\CIRCLE\ar@{-}[d]\ar@{-}[dl]\ar@{-}[drrrr] &&&&\\
\CIRCLE\ar@{-}[r]\ar@/_1pc/@{-}[rr]\ar@/_2pc/@{-}[rrr]\ar@/_3pc/@{-}[rrrr]& \CIRCLE\ar@{-}[r]\ar@/_1pc/@{-}[rr]\ar@/_2pc/@{-}[rrr]& \CIRCLE\ar@{-}[r]\ar@/_1pc/@{-}[rr]\ar@/_2pc/@{-}[rrr]& \CIRCLE\ar@{-}[r]\ar@/_1pc/@{-}[rr]& \CIRCLE\ar@{-}[r]& \color{blue}\CIRCLE\\
&&&&&\\}$$
Removing the two degree-$4$ nodes and that degree-$5$ node will reduce it to a $(5,10)$ absorption sets.
\end{enumerate}
      \end{enumerate}
    \end{enumerate}
    \item $[6,6,6,5,5,4,4,4]$: There should be no more than two intrinsic connections among the
three degree-$4$ nodes. Otherwise, the group of degree-$4$ nodes will not match the group of the other five nodes remained
in the set.
    \begin{enumerate}\setlength{\itemsep}{0pt}
      \item No connection as shown in \figurename \ref{figinternal3:sub:a}: removing any degree-$4$ node will reduce it to a
$(7,10)$ absorption set.
      \item One connection as shown in \figurename \ref{figinternal3:sub:b}: removing the topmost degree-$4$ node
will reduce it to a $(7,10)$ absorption set.
      \item Two connections: Now, there are eight edges coming out the
group of degree-$4$ nodes as shown in \figurename \ref{figinternal3:sub:c}. However, each of the two degree-$5$ nodes
and each of the three degree-$6$ nodes needs one and two connections emanating
from the group of degree-$4$ nodes, respectively. That makes eight.
Thus, removing the three degree-$4$ nodes will reduce it to a $(5,10)$
absorption set.
    \end{enumerate}
    \item $[6,6,6,6,4,4,4,4]$: The group of degree-$6$ nodes need at
    least twelve edges from the group of degree-$4$ nodes. So there should be no more than two connections among the four degree-$4$
      nodes.
    \begin{enumerate}\setlength{\itemsep}{0pt}
      \item No connection: removing any degree-$4$ node in \figurename \ref{figinternal4:sub:a} will reduce it to a
$(7,10)$ absorption set.
      \item One connection: removing either topmost degree-$4$ node in \figurename \ref{figinternal4:sub:b} will
reduce it to a $(7,10)$ absorption set.
      \item Two connections: there are two cases:
      \begin{enumerate}\setlength{\itemsep}{0pt}
        \item Removing the bottom-right degree-$4$ node in \figurename \ref{figinternal4:sub:c} will
reduce it to a $(7,10)$ absorption set.
        \item Removing either the top or the bottom couple of
degree-$4$ nodes in \figurename \ref{figinternal4:sub:d} will reduce it to a $(6,10)$ absorption set.
      \end{enumerate}
    \end{enumerate}
  \end{enumerate}

\bigskip

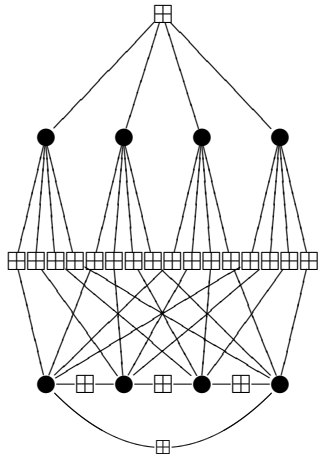
\begin{figure}[!t]
\centering
$\xymatrix@M=0pt@W=0pt@R=40pt@C=-4.7pt
{
&&&&&&&&&&&&&&&\boxplus\ar@{-}[dllll]\ar@{-}[dllllllllllll]\ar@{-}[drrrr]\ar@{-}[drrrrrrrrrrrr]&&&&&&&&&&&&&&&\\
&&& \CIRCLE\ar@{-}[dl]\ar@{-}[dlll]\ar@{-}[dr]\ar@{-}[drrr]&&&&&&&& \CIRCLE\ar@{-}[dl]\ar@{-}[dlll]\ar@{-}[dr]\ar@{-}[drrr]&&&&&&&& \CIRCLE\ar@{-}[dl]\ar@{-}[dlll]\ar@{-}[dr]\ar@{-}[drrr]&&&&&&&& \CIRCLE\ar@{-}[dl]\ar@{-}[dlll]\ar@{-}[dr]\ar@{-}[drrr]&&&\\
\boxplus&\hspace{9pt}& \boxplus&\hspace{9pt}& \boxplus&\hspace{9pt}& \boxplus&\hspace{9pt}& \boxplus&\hspace{9pt}& \boxplus&\hspace{9pt}& \boxplus&\hspace{9pt}& \boxplus&\hspace{9pt}& \boxplus&\hspace{9pt}& \boxplus&\hspace{9pt}& \boxplus&\hspace{9pt}& \boxplus&\hspace{9pt}& \boxplus&\hspace{9pt}& \boxplus&\hspace{9pt}& \boxplus&\hspace{9pt}& \boxplus\\
&&& \CIRCLE\ar@/_2pc/@{-}[rrrrrrrrrrrrrrrrrrrrrrrr]|-{\boxplus}\ar@{-}[rrrr]\ar@{-}[ulll]\ar@{-}[urrrrr]\ar@{-}[urrrrrrrrrrrrr]\ar@{-}[urrrrrrrrrrrrrrrrrrrrr]&&&& \boxplus&&&& \CIRCLE\ar@{-}[llll]\ar@{-}[rrrr]\ar@{-}[ul]\ar@{-}[ulllllllll]\ar@{-}[urrrrrrr]\ar@{-}[urrrrrrrrrrrrrrr]&&&& \boxplus&&&& \CIRCLE\ar@{-}[llll]\ar@{-}[rrrr]\ar@{-}[ulllllllllllllll]\ar@{-}[ulllllll]\ar@{-}[ur]\ar@{-}[urrrrrrrrr]&&&& \boxplus&&&& \CIRCLE\ar@{-}[llll]\ar@{-}[ulllllllllllllllllllll]\ar@{-}[ulllllllllllll]\ar@{-}[ulllll]\ar@{-}[urrr]&&&\\
}$
\caption{One check node connecting to $(8,4)$ sets four times.}
\label{figa=8b484}
\end{figure}

If a neighboring check node connecting to the set four times is considered, then the smallest such size-$8$ absorption set would be $(8,4)$, \figurename \ref{figa=8b484}. Removing any two degree-$5$ nodes will reduce it to a $(6,10)$ absorption set. To obtain such $(8,6)$ sets, let us remove one edge from \figurename \ref{figa=8b484}.

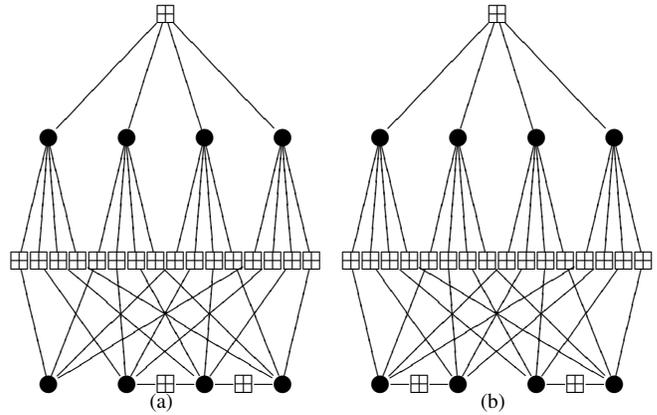
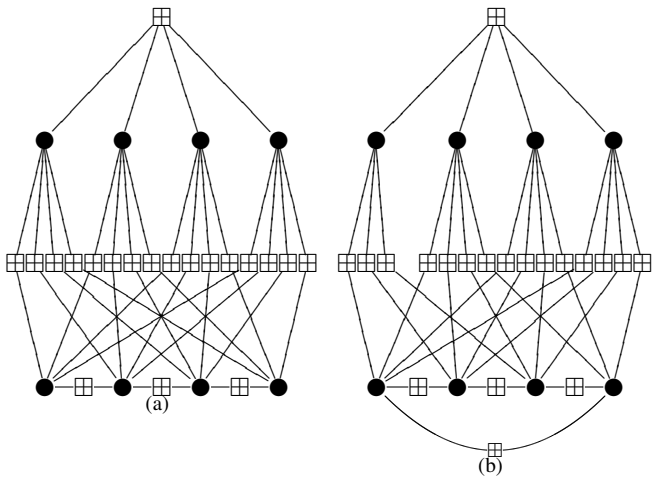
\begin{figure}[!t]
\subfigure[] 
{
    \label{figa=8b486:sub:a}
$\xymatrix@M=0pt@W=0pt@R=40pt@C=-4.7pt
{
&&&&&&&&&&&&&&&\boxplus\ar@{-}[dllll]\ar@{-}[dllllllllllll]\ar@{-}[drrrr]\ar@{-}[drrrrrrrrrrrr]&&&&&&&&&&&&&&&\\
&&& \CIRCLE\ar@{-}[dl]\ar@{-}[dlll]\ar@{-}[dr]\ar@{-}[drrr]&&&&&&&& \CIRCLE\ar@{-}[dl]\ar@{-}[dlll]\ar@{-}[dr]\ar@{-}[drrr]&&&&&&&& \CIRCLE\ar@{-}[dl]\ar@{-}[dlll]\ar@{-}[dr]\ar@{-}[drrr]&&&&&&&& \CIRCLE\ar@{-}[dl]\ar@{-}[dlll]\ar@{-}[dr]\ar@{-}[drrr]&&&\\
\boxplus&\hspace{9pt}& \boxplus&\hspace{9pt}& \boxplus&\hspace{9pt}& \boxplus&\hspace{9pt}& \boxplus&\hspace{9pt}& \boxplus&\hspace{9pt}& \boxplus&\hspace{9pt}& \boxplus&\hspace{9pt}& \boxplus&\hspace{9pt}& \boxplus&\hspace{9pt}& \boxplus&\hspace{9pt}& \boxplus&\hspace{9pt}& \boxplus&\hspace{9pt}& \boxplus&\hspace{9pt}& \boxplus&\hspace{9pt}& \boxplus\\
&&& \CIRCLE\ar@{-}[rrrr]\ar@{-}[ulll]\ar@{-}[urrrrr]\ar@{-}[urrrrrrrrrrrrr]\ar@{-}[urrrrrrrrrrrrrrrrrrrrr]&&&& \boxplus&&&& \CIRCLE\ar@{-}[llll]\ar@{-}[rrrr]\ar@{-}[ul]\ar@{-}[ulllllllll]\ar@{-}[urrrrrrr]\ar@{-}[urrrrrrrrrrrrrrr]&&&& \boxplus&&&& \CIRCLE\ar@{-}[llll]\ar@{-}[rrrr]\ar@{-}[ulllllllllllllll]\ar@{-}[ulllllll]\ar@{-}[ur]\ar@{-}[urrrrrrrrr]&&&& \boxplus&&&& \CIRCLE\ar@{-}[llll]\ar@{-}[ulllllllllllllllllllll]\ar@{-}[ulllllllllllll]\ar@{-}[ulllll]\ar@{-}[urrr]&&&\\
}$ }
\subfigure[] 
{
    \label{figa=8b486:sub:b}
$\xymatrix@M=0pt@W=0pt@R=40pt@C=-4.7pt
{
&&&&&&&&&&&&&&&\boxplus\ar@{-}[dllll]\ar@{-}[dllllllllllll]\ar@{-}[drrrr]\ar@{-}[drrrrrrrrrrrr]&&&&&&&&&&&&&&&\\
&&& \CIRCLE\ar@{-}[dl]\ar@{-}[dlll]\ar@{-}[dr]&&&&&&&& \CIRCLE\ar@{-}[dl]\ar@{-}[dlll]\ar@{-}[dr]\ar@{-}[drrr]&&&&&&&& \CIRCLE\ar@{-}[dl]\ar@{-}[dlll]\ar@{-}[dr]\ar@{-}[drrr]&&&&&&&& \CIRCLE\ar@{-}[dl]\ar@{-}[dlll]\ar@{-}[dr]\ar@{-}[drrr]&&&\\
\boxplus&\hspace{9pt}& \boxplus&\hspace{9pt}& \boxplus&\hspace{9pt}& \hspace{9pt}&\hspace{9pt}& \boxplus&\hspace{9pt}& \boxplus&\hspace{9pt}& \boxplus&\hspace{9pt}& \boxplus&\hspace{9pt}& \boxplus&\hspace{9pt}& \boxplus&\hspace{9pt}& \boxplus&\hspace{9pt}& \boxplus&\hspace{9pt}& \boxplus&\hspace{9pt}& \boxplus&\hspace{9pt}& \boxplus&\hspace{9pt}& \boxplus\\
&&& \CIRCLE\ar@/_2pc/@{-}[rrrrrrrrrrrrrrrrrrrrrrrr]|-{\boxplus}\ar@{-}[rrrr]\ar@{-}[ulll]\ar@{-}[urrrrr]\ar@{-}[urrrrrrrrrrrrr]\ar@{-}[urrrrrrrrrrrrrrrrrrrrr]&&&& \boxplus&&&& \CIRCLE\ar@{-}[llll]\ar@{-}[rrrr]\ar@{-}[ul]\ar@{-}[ulllllllll]\ar@{-}[urrrrrrr]\ar@{-}[urrrrrrrrrrrrrrr]&&&& \boxplus&&&& \CIRCLE\ar@{-}[llll]\ar@{-}[rrrr]\ar@{-}[ulllllllllllllll]\ar@{-}[ulllllll]\ar@{-}[ur]\ar@{-}[urrrrrrrrr]&&&& \boxplus&&&& \CIRCLE\ar@{-}[llll]\ar@{-}[ulllllllllllll]\ar@{-}[ulllll]\ar@{-}[urrr]&&&\\
}$ }
 \caption{One check node connecting to $(8,6)$ sets four times.}
\label{figa=8b486:sub} 
\end{figure}

By removing either the bottom-left or the bottom-right degree-$5$ node from \figurename \ref{figa=8b486:sub:a}, or the degree-$4$ node from \figurename \ref{figa=8b486:sub:b}, respectively, we obtain a $(7,10)$ absorption set.

\begin{figure}[!t]
\subfigure[] 
{
    \label{figa=8b488:sub:a}
$\xymatrix@M=0pt@W=0pt@R=40pt@C=-4.7pt
{
&&&&&&&&&&&&&&&\boxplus\ar@{-}[dllll]\ar@{-}[dllllllllllll]\ar@{-}[drrrr]\ar@{-}[drrrrrrrrrrrr]&&&&&&&&&&&&&&&\\
&&& \CIRCLE\ar@{-}[dl]\ar@{-}[dlll]\ar@{-}[dr]\ar@{-}[drrr]&&&&&&&& \CIRCLE\ar@{-}[dl]\ar@{-}[dlll]\ar@{-}[dr]\ar@{-}[drrr]&&&&&&&& \CIRCLE\ar@{-}[dl]\ar@{-}[dlll]\ar@{-}[dr]\ar@{-}[drrr]&&&&&&&& \CIRCLE\ar@{-}[dl]\ar@{-}[dlll]\ar@{-}[dr]\ar@{-}[drrr]&&&\\
\boxplus&\hspace{9pt}& \boxplus&\hspace{9pt}& \boxplus&\hspace{9pt}& \boxplus&\hspace{9pt}& \boxplus&\hspace{9pt}& \boxplus&\hspace{9pt}& \boxplus&\hspace{9pt}& \boxplus&\hspace{9pt}& \boxplus&\hspace{9pt}& \boxplus&\hspace{9pt}& \boxplus&\hspace{9pt}& \boxplus&\hspace{9pt}& \boxplus&\hspace{9pt}& \boxplus&\hspace{9pt}& \boxplus&\hspace{9pt}& \boxplus\\
&&& \CIRCLE\ar@{-}[ulll]\ar@{-}[urrrrr]\ar@{-}[urrrrrrrrrrrrr]\ar@{-}[urrrrrrrrrrrrrrrrrrrrr]&&&&&&&& \CIRCLE\ar@{-}[rrrr]\ar@{-}[ul]\ar@{-}[ulllllllll]\ar@{-}[urrrrrrr]\ar@{-}[urrrrrrrrrrrrrrr]&&&& \boxplus&&&& \CIRCLE\ar@{-}[llll]\ar@{-}[rrrr]\ar@{-}[ulllllllllllllll]\ar@{-}[ulllllll]\ar@{-}[ur]\ar@{-}[urrrrrrrrr]&&&& \boxplus&&&& \CIRCLE\ar@{-}[llll]\ar@{-}[ulllllllllllllllllllll]\ar@{-}[ulllllllllllll]\ar@{-}[ulllll]\ar@{-}[urrr]&&&\\
}$ }
\subfigure[] 
{
    \label{figa=8b488:sub:b}
$\xymatrix@M=0pt@W=0pt@R=40pt@C=-4.7pt
{
&&&&&&&&&&&&&&&\boxplus\ar@{-}[dllll]\ar@{-}[dllllllllllll]\ar@{-}[drrrr]\ar@{-}[drrrrrrrrrrrr]&&&&&&&&&&&&&&&\\
&&& \CIRCLE\ar@{-}[dl]\ar@{-}[dlll]\ar@{-}[dr]\ar@{-}[drrr]&&&&&&&& \CIRCLE\ar@{-}[dl]\ar@{-}[dlll]\ar@{-}[dr]\ar@{-}[drrr]&&&&&&&& \CIRCLE\ar@{-}[dl]\ar@{-}[dlll]\ar@{-}[dr]\ar@{-}[drrr]&&&&&&&& \CIRCLE\ar@{-}[dl]\ar@{-}[dlll]\ar@{-}[dr]\ar@{-}[drrr]&&&\\
\boxplus&\hspace{9pt}& \boxplus&\hspace{9pt}& \boxplus&\hspace{9pt}& \boxplus&\hspace{9pt}& \boxplus&\hspace{9pt}& \boxplus&\hspace{9pt}& \boxplus&\hspace{9pt}& \boxplus&\hspace{9pt}& \boxplus&\hspace{9pt}& \boxplus&\hspace{9pt}& \boxplus&\hspace{9pt}& \boxplus&\hspace{9pt}& \boxplus&\hspace{9pt}& \boxplus&\hspace{9pt}& \boxplus&\hspace{9pt}& \boxplus\\
&&& \CIRCLE\ar@{-}[rrrr]\ar@{-}[ulll]\ar@{-}[urrrrr]\ar@{-}[urrrrrrrrrrrrr]\ar@{-}[urrrrrrrrrrrrrrrrrrrrr]&&&& \boxplus&&&& \CIRCLE\ar@{-}[llll]\ar@{-}[ul]\ar@{-}[ulllllllll]\ar@{-}[urrrrrrr]\ar@{-}[urrrrrrrrrrrrrrr]&&&&&&&& \CIRCLE\ar@{-}[rrrr]\ar@{-}[ulllllllllllllll]\ar@{-}[ulllllll]\ar@{-}[ur]\ar@{-}[urrrrrrrrr]&&&& \boxplus&&&& \CIRCLE\ar@{-}[llll]\ar@{-}[ulllllllllllllllllllll]\ar@{-}[ulllllllllllll]\ar@{-}[ulllll]\ar@{-}[urrr]&&&\\
}$ }\\
\subfigure[] 
{
    \label{figa=8b488:sub:c}
$\xymatrix@M=0pt@W=0pt@R=40pt@C=-4.7pt
{
&&&&&&&&&&&&&&&\boxplus\ar@{-}[dllll]\ar@{-}[dllllllllllll]\ar@{-}[drrrr]\ar@{-}[drrrrrrrrrrrr]&&&&&&&&&&&&&&&\\
&&& \CIRCLE\ar@{-}[dl]\ar@{-}[dr]\ar@{-}[drrr]&&&&&&&& \CIRCLE\ar@{-}[dl]\ar@{-}[dlll]\ar@{-}[dr]\ar@{-}[drrr]&&&&&&&& \CIRCLE\ar@{-}[dl]\ar@{-}[dlll]\ar@{-}[dr]\ar@{-}[drrr]&&&&&&&& \CIRCLE\ar@{-}[dl]\ar@{-}[dlll]\ar@{-}[dr]\ar@{-}[drrr]&&&\\
\hspace{9pt}&\hspace{9pt}& \boxplus&\hspace{9pt}& \boxplus&\hspace{9pt}& \boxplus&\hspace{9pt}& \boxplus&\hspace{9pt}& \boxplus&\hspace{9pt}& \boxplus&\hspace{9pt}& \boxplus&\hspace{9pt}& \boxplus&\hspace{9pt}& \boxplus&\hspace{9pt}& \boxplus&\hspace{9pt}& \boxplus&\hspace{9pt}& \boxplus&\hspace{9pt}& \boxplus&\hspace{9pt}& \boxplus&\hspace{9pt}& \boxplus\\
&&& \CIRCLE\ar@{-}[rrrr]\ar@{-}[urrrrr]\ar@{-}[urrrrrrrrrrrrr]\ar@{-}[urrrrrrrrrrrrrrrrrrrrr]&&&& \boxplus&&&& \CIRCLE\ar@{-}[llll]\ar@{-}[rrrr]\ar@{-}[ul]\ar@{-}[ulllllllll]\ar@{-}[urrrrrrr]\ar@{-}[urrrrrrrrrrrrrrr]&&&& \boxplus&&&& \CIRCLE\ar@{-}[llll]\ar@{-}[rrrr]\ar@{-}[ulllllllllllllll]\ar@{-}[ulllllll]\ar@{-}[ur]\ar@{-}[urrrrrrrrr]&&&& \boxplus&&&& \CIRCLE\ar@{-}[llll]\ar@{-}[ulllllllllllllllllllll]\ar@{-}[ulllllllllllll]\ar@{-}[ulllll]\ar@{-}[urrr]&&&\\
}$ }
\subfigure[] 
{
    \label{figa=8b488:sub:d}
$\xymatrix@M=0pt@W=0pt@R=40pt@C=-4.7pt
{
&&&&&&&&&&&&&&&\boxplus\ar@{-}[dllll]\ar@{-}[dllllllllllll]\ar@{-}[drrrr]\ar@{-}[drrrrrrrrrrrr]&&&&&&&&&&&&&&&\\
&&& \CIRCLE\ar@{-}[dlll]\ar@{-}[dr]\ar@{-}[drrr]&&&&&&&& \CIRCLE\ar@{-}[dl]\ar@{-}[dlll]\ar@{-}[dr]\ar@{-}[drrr]&&&&&&&& \CIRCLE\ar@{-}[dl]\ar@{-}[dlll]\ar@{-}[dr]\ar@{-}[drrr]&&&&&&&& \CIRCLE\ar@{-}[dl]\ar@{-}[dlll]\ar@{-}[dr]\ar@{-}[drrr]&&&\\
\boxplus&\hspace{9pt}& \hspace{9pt}&\hspace{9pt}& \boxplus&\hspace{9pt}& \boxplus&\hspace{9pt}& \boxplus&\hspace{9pt}& \boxplus&\hspace{9pt}& \boxplus&\hspace{9pt}& \boxplus&\hspace{9pt}& \boxplus&\hspace{9pt}& \boxplus&\hspace{9pt}& \boxplus&\hspace{9pt}& \boxplus&\hspace{9pt}& \boxplus&\hspace{9pt}& \boxplus&\hspace{9pt}& \boxplus&\hspace{9pt}& \boxplus\\
&&& \CIRCLE\ar@{-}[rrrr]\ar@{-}[ulll]\ar@{-}[urrrrr]\ar@{-}[urrrrrrrrrrrrr]\ar@{-}[urrrrrrrrrrrrrrrrrrrrr]&&&& \boxplus&&&& \CIRCLE\ar@{-}[llll]\ar@{-}[rrrr]\ar@{-}[ul]\ar@{-}[urrrrrrr]\ar@{-}[urrrrrrrrrrrrrrr]&&&& \boxplus&&&& \CIRCLE\ar@{-}[llll]\ar@{-}[rrrr]\ar@{-}[ulllllllllllllll]\ar@{-}[ulllllll]\ar@{-}[ur]\ar@{-}[urrrrrrrrr]&&&& \boxplus&&&& \CIRCLE\ar@{-}[llll]\ar@{-}[ulllllllllllllllllllll]\ar@{-}[ulllllllllllll]\ar@{-}[ulllll]\ar@{-}[urrr]&&&\\
}$ }\\
\subfigure[] 
{
    \label{figa=8b488:sub:e}
$\xymatrix@M=0pt@W=0pt@R=40pt@C=-4.7pt
{
&&&&&&&&&&&&&&&\boxplus\ar@{-}[dllll]\ar@{-}[dllllllllllll]\ar@{-}[drrrr]\ar@{-}[drrrrrrrrrrrr]&&&&&&&&&&&&&&&\\
&&& \CIRCLE\ar@{-}[dl]\ar@{-}[dlll]\ar@{-}[dr]&&&&&&&& \CIRCLE\ar@{-}[dl]\ar@{-}[dlll]\ar@{-}[dr]\ar@{-}[drrr]&&&&&&&& \CIRCLE\ar@{-}[dl]\ar@{-}[dlll]\ar@{-}[dr]\ar@{-}[drrr]&&&&&&&& \CIRCLE\ar@{-}[dl]\ar@{-}[dlll]\ar@{-}[dr]&&&\\
\boxplus&\hspace{9pt}& \boxplus&\hspace{9pt}& \boxplus&\hspace{9pt}& \hspace{9pt}&\hspace{9pt}& \boxplus&\hspace{9pt}& \boxplus&\hspace{9pt}& \boxplus&\hspace{9pt}& \boxplus&\hspace{9pt}& \boxplus&\hspace{9pt}& \boxplus&\hspace{9pt}& \boxplus&\hspace{9pt}& \boxplus&\hspace{9pt}& \boxplus&\hspace{9pt}& \boxplus&\hspace{9pt}& \boxplus&\hspace{9pt}& \hspace{9pt}\\
&&& \CIRCLE\ar@/_2pc/@{-}[rrrrrrrrrrrrrrrrrrrrrrrr]|-{\boxplus}\ar@{-}[rrrr]\ar@{-}[ulll]\ar@{-}[urrrrr]\ar@{-}[urrrrrrrrrrrrr]\ar@{-}[urrrrrrrrrrrrrrrrrrrrr]&&&& \boxplus&&&& \CIRCLE\ar@{-}[llll]\ar@{-}[rrrr]\ar@{-}[ul]\ar@{-}[ulllllllll]\ar@{-}[urrrrrrr]\ar@{-}[urrrrrrrrrrrrrrr]&&&& \boxplus&&&& \CIRCLE\ar@{-}[llll]\ar@{-}[rrrr]\ar@{-}[ulllllllllllllll]\ar@{-}[ulllllll]\ar@{-}[ur]\ar@{-}[urrrrrrrrr]&&&& \boxplus&&&& \CIRCLE\ar@{-}[llll]\ar@{-}[ulllllllllllll]\ar@{-}[ulllll]&&&\\
}$ }
\subfigure[] 
{
    \label{figa=8b488:sub:f}
$\xymatrix@M=0pt@W=0pt@R=40pt@C=-4.7pt
{
&&&&&&&&&&&&&&&\boxplus\ar@{-}[dllll]\ar@{-}[dllllllllllll]\ar@{-}[drrrr]\ar@{-}[drrrrrrrrrrrr]&&&&&&&&&&&&&&&\\
&&& \CIRCLE\ar@{-}[dl]\ar@{-}[dlll]\ar@{-}[dr]&&&&&&&& \CIRCLE\ar@{-}[dl]\ar@{-}[dlll]\ar@{-}[dr]\ar@{-}[drrr]&&&&&&&& \CIRCLE\ar@{-}[dl]\ar@{-}[dlll]\ar@{-}[dr]\ar@{-}[drrr]&&&&&&&& \CIRCLE\ar@{-}[dl]\ar@{-}[dr]\ar@{-}[drrr]&&&\\
\boxplus&\hspace{9pt}& \boxplus&\hspace{9pt}& \boxplus&\hspace{9pt}& \hspace{9pt}&\hspace{9pt}& \boxplus&\hspace{9pt}& \boxplus&\hspace{9pt}& \boxplus&\hspace{9pt}& \boxplus&\hspace{9pt}& \boxplus&\hspace{9pt}& \boxplus&\hspace{9pt}& \boxplus&\hspace{9pt}& \boxplus&\hspace{9pt}& \hspace{9pt}&\hspace{9pt}& \boxplus&\hspace{9pt}& \boxplus&\hspace{9pt}& \boxplus\\
&&& \CIRCLE\ar@/_2pc/@{-}[rrrrrrrrrrrrrrrrrrrrrrrr]|-{\boxplus}\ar@{-}[rrrr]\ar@{-}[ulll]\ar@{-}[urrrrr]\ar@{-}[urrrrrrrrrrrrr]&&&& \boxplus&&&& \CIRCLE\ar@{-}[llll]\ar@{-}[rrrr]\ar@{-}[ul]\ar@{-}[ulllllllll]\ar@{-}[urrrrrrr]\ar@{-}[urrrrrrrrrrrrrrr]&&&& \boxplus&&&& \CIRCLE\ar@{-}[llll]\ar@{-}[rrrr]\ar@{-}[ulllllllllllllll]\ar@{-}[ulllllll]\ar@{-}[ur]\ar@{-}[urrrrrrrrr]&&&& \boxplus&&&& \CIRCLE\ar@{-}[llll]\ar@{-}[ulllllllllllll]\ar@{-}[ulllll]\ar@{-}[urrr]&&&\\
}$ }\caption{One check node connecting to $(8,8)$ sets four times.}
\label{figa=8b488:sub} 
\end{figure}
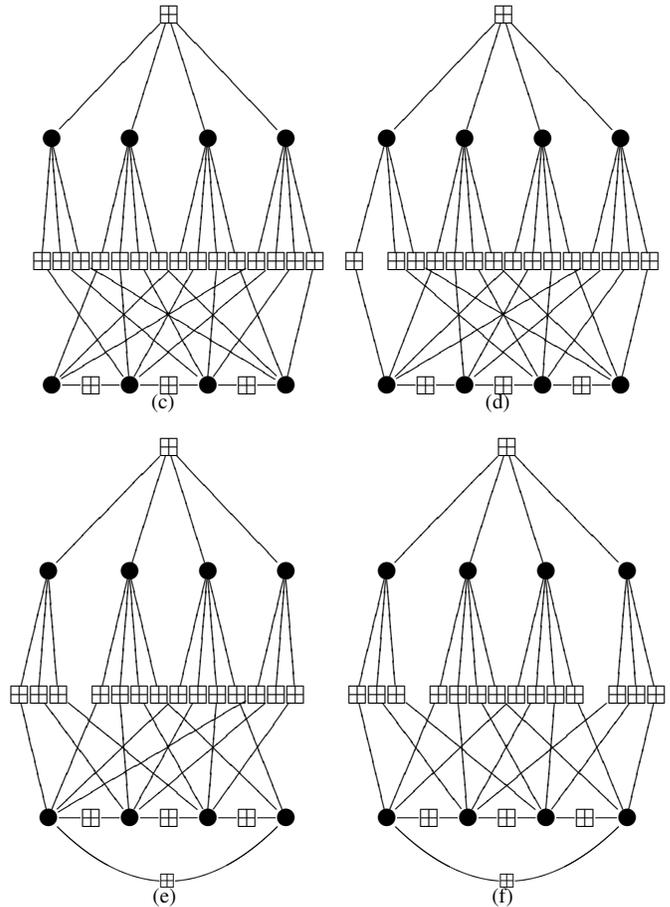

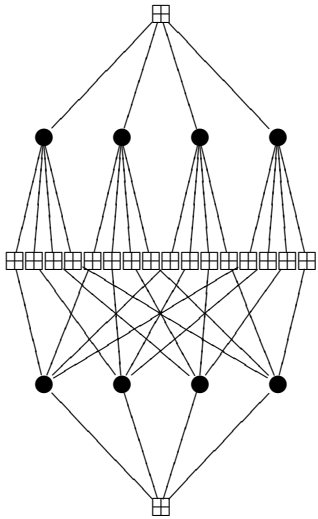
\begin{figure}[!t]
\centering
$\xymatrix@M=0pt@W=0pt@R=40pt@C=-4.7pt
{
&&&&&&&&&&&&&&&\boxplus\ar@{-}[dllll]\ar@{-}[dllllllllllll]\ar@{-}[drrrr]\ar@{-}[drrrrrrrrrrrr]&&&&&&&&&&&&&&&\\
&&& \CIRCLE\ar@{-}[dl]\ar@{-}[dlll]\ar@{-}[dr]\ar@{-}[drrr]&&&&&&&& \CIRCLE\ar@{-}[dl]\ar@{-}[dlll]\ar@{-}[dr]\ar@{-}[drrr]&&&&&&&& \CIRCLE\ar@{-}[dl]\ar@{-}[dlll]\ar@{-}[dr]\ar@{-}[drrr]&&&&&&&& \CIRCLE\ar@{-}[dl]\ar@{-}[dlll]\ar@{-}[dr]\ar@{-}[drrr]&&&\\
\boxplus&\hspace{9pt}& \boxplus&\hspace{9pt}& \boxplus&\hspace{9pt}& \boxplus&\hspace{9pt}& \boxplus&\hspace{9pt}& \boxplus&\hspace{9pt}& \boxplus&\hspace{9pt}& \boxplus&\hspace{9pt}& \boxplus&\hspace{9pt}& \boxplus&\hspace{9pt}& \boxplus&\hspace{9pt}& \boxplus&\hspace{9pt}& \boxplus&\hspace{9pt}& \boxplus&\hspace{9pt}& \boxplus&\hspace{9pt}& \boxplus\\
&&& \CIRCLE\ar@{-}[ulll]\ar@{-}[urrrrr]\ar@{-}[urrrrrrrrrrrrr]\ar@{-}[urrrrrrrrrrrrrrrrrrrrr]&&&&&&&& \CIRCLE\ar@{-}[ul]\ar@{-}[ulllllllll]\ar@{-}[urrrrrrr]\ar@{-}[urrrrrrrrrrrrrrr]&&&&&&&& \CIRCLE\ar@{-}[ulllllllllllllll]\ar@{-}[ulllllll]\ar@{-}[ur]\ar@{-}[urrrrrrrrr]&&&&&&&& \CIRCLE\ar@{-}[ulllllllllllllllllllll]\ar@{-}[ulllllllllllll]\ar@{-}[ulllll]\ar@{-}[urrr]&&&\\
&&&&&&&&&&&&&&& \boxplus\ar@{-}[ullll]\ar@{-}[ullllllllllll]\ar@{-}[urrrr]\ar@{-}[urrrrrrrrrrrr]&&&&&&&&&&&&&&&\\
}$\caption{Two check nodes connecting to $(8,8)$ set four times.}
\label{figa=8b488:sub:g} 
\end{figure}

Then to obtain such $(8,8)$ absorption sets, first we remove one edge from \figurename \ref{figa=8b486:sub}, which gives us \figurename \ref{figa=8b488:sub}. In addition, the possible topology with two neighboring check nodes connecting to the set four times is shown in \figurename \ref{figa=8b488:sub:g}. Removing the degree-$4$ node, the bottom-left degree-$4$ node or the bottom-right degree-$4$ node in \figurename \ref{figa=8b488:sub:a}, \ref{figa=8b488:sub:c} or \ref{figa=8b488:sub:e}, respectively, reduces it to a $(7,10)$ absorption set, while removing the two degree-$4$ nodes in \figurename \ref{figa=8b488:sub:f} reduces it to a $(6,12)$ absorption set.

Note that \figurename \ref{figa=8b488:sub:b} and \ref{figa=8b488:sub:g} are in the class $[5,5,5,5,5,5,5,5]$, so after searching $\mathbf{H}$ with \figurename \ref{figa=8b488:sub:d}, Lemma \ref{lema=8} results.

\section{Proof of Claim \ref{claima=8}}\label{proofclaima=8}

If a check node connecting to the set an even number, but more than twice is allowed, we obtain \figurename \ref{figa=8b488:sub:b} and \ref{figa=8b488:sub:g}. Let us find the other three by restricting that a satisfied check node can only be connected to the set twice.

We start with one node: $$\xymatrix@M=0pt@W=0pt@R=15.78pt@C=15.78pt
{
& \color{blue}\CIRCLE\ar@{-}[r]\ar@{-}[dl]\ar@{-}[ddl]\ar@{-}[drr]\ar@{-}[ddrr]& \CIRCLE&\\
\CIRCLE &&& \CIRCLE\\
\CIRCLE&&& \CIRCLE\\
& \CIRCLE& \CIRCLE&\\
}$$
As Step 2, if the bottom two nodes are

\subsection{not connected}
\begin{figure}[!t]
\centering \subfigure[] 
{
    \label{fig88a:sub:a}
$\xymatrix@M=0pt@W=0pt@R=15.78pt@C=15.78pt
{
& \color{blue}\CIRCLE\ar@{-}[r]\ar@{-}[dl]\ar@{-}[ddl]\ar@{-}[drr]\ar@{-}[ddrr]& \CIRCLE&\\
\CIRCLE &&& \CIRCLE\\
\CIRCLE&&& \CIRCLE\\
& \color{blue}\CIRCLE\ar@{-}[ul]\ar@{-}[uul]\ar@{-}[uuur]\ar@{-}[uurr]\ar@{-}[urr]& \color{blue}\CIRCLE\ar@{-}[ull]\ar@{-}[uull]\ar@{-}[uuu]\ar@{-}[uur]\ar@{-}[ur]&\\
}$ }\quad
\subfigure[] 
{
    \label{fig88a:sub:b}
$\xymatrix@M=0pt@W=0pt@R=15.78pt@C=15.78pt
{
& \color{blue}\CIRCLE\ar@{-}[r]\ar@{-}[dl]\ar@{-}[ddl]\ar@{-}[drr]\ar@{-}[ddrr]&\color{blue} \CIRCLE\ar@{-}[dll]\ar@{-}[dr]&\\
\color{blue}\CIRCLE\ar@{-}[d] &&& \color{blue}\CIRCLE\ar@{-}[d]\\
\color{blue}\CIRCLE\ar@{-}[rrr]&&& \color{blue}\CIRCLE\\
& \color{blue}\CIRCLE\ar@{-}[ul]\ar@{-}[uul]\ar@{-}[uuur]\ar@{-}[uurr]\ar@{-}[urr]& \color{blue}\CIRCLE\ar@{-}[ull]\ar@{-}[uull]\ar@{-}[uuu]\ar@{-}[uur]\ar@{-}[ur]&\\
}$ } \caption{First possible topology of $(8,8)$ absorption sets.}
\label{fig88a:sub} 
\end{figure}
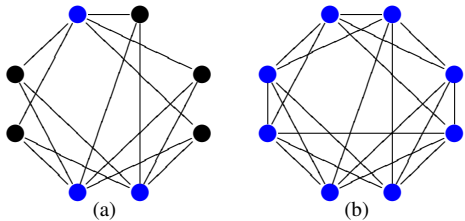

We must have \figurename \ref{fig88a:sub:a}.

After connecting the remaining five nodes we obtain \figurename \ref{fig88a:sub:b}.
Reorganize \figurename \ref{fig88a:sub:b} into a symmetric form as \figurename \ref{fig12:sub:a}.

\subsection{connected}
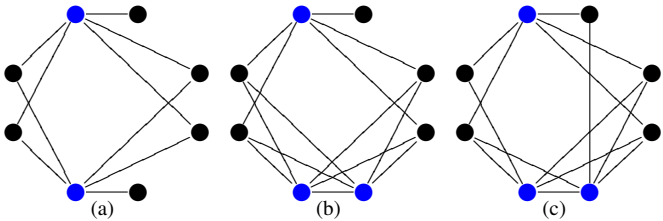
\begin{figure}[!t]
\centering \subfigure[] 
{
    \label{fig88b:sub:a}
$\xymatrix@M=0pt@W=0pt@R=15.78pt@C=15.78pt
{
& \color{blue}\CIRCLE\ar@{-}[r]\ar@{-}[dl]\ar@{-}[ddl]\ar@{-}[drr]\ar@{-}[ddrr]& \CIRCLE&\\
\CIRCLE &&& \CIRCLE\\
\CIRCLE&&& \CIRCLE\\
& \color{blue}\CIRCLE\ar@{-}[r]\ar@{-}[ul]\ar@{-}[uul]\ar@{-}[uurr]\ar@{-}[urr]& \CIRCLE&\\
}$ }
\subfigure[] 
{
    \label{fig88b:sub:b}
$\xymatrix@M=0pt@W=0pt@R=15.78pt@C=15.78pt
{
& \color{blue}\CIRCLE\ar@{-}[r]\ar@{-}[dl]\ar@{-}[ddl]\ar@{-}[drr]\ar@{-}[ddrr]& \CIRCLE&\\
\CIRCLE &&& \CIRCLE\\
\CIRCLE&&& \CIRCLE\\
& \color{blue}\CIRCLE\ar@{-}[r]\ar@{-}[ul]\ar@{-}[uul]\ar@{-}[uurr]\ar@{-}[urr]& \color{blue}\CIRCLE\ar@{-}[ur]\ar@{-}[uur]\ar@{-}[uull]\ar@{-}[ull]&\\
}$ }
\subfigure[] 
{
    \label{fig88b:sub:c}
$\xymatrix@M=0pt@W=0pt@R=15.78pt@C=15.78pt
{
& \color{blue}\CIRCLE\ar@{-}[r]\ar@{-}[dl]\ar@{-}[ddl]\ar@{-}[drr]\ar@{-}[ddrr]& \CIRCLE&\\
\CIRCLE &&& \CIRCLE\\
\CIRCLE&&& \CIRCLE\\
& \color{blue}\CIRCLE\ar@{-}[r]\ar@{-}[ul]\ar@{-}[uul]\ar@{-}[uurr]\ar@{-}[urr]& \color{blue}\CIRCLE\ar@{-}[uuu]\ar@{-}[ur]\ar@{-}[uur]\ar@{-}[ull]&\\
}$ } \caption{Finding possible topology of $(8,8)$ absorption sets -- Step 3.}
\label{fig88b:sub} 
\end{figure}

There is only one choice for either one of them as shown in \figurename \ref{fig88b:sub:a}.

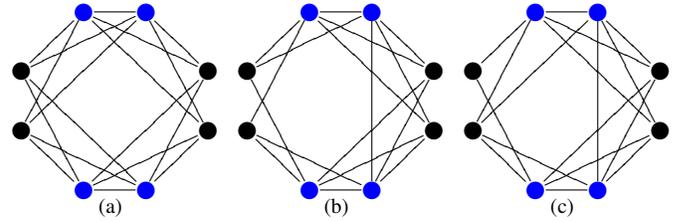
\begin{figure}[!t]
\centering \subfigure[] 
{
    \label{fig88c:sub:a}
$\xymatrix@M=0pt@W=0pt@R=15.78pt@C=15.78pt
{
& \color{blue}\CIRCLE\ar@{-}[r]\ar@{-}[dl]\ar@{-}[ddl]\ar@{-}[drr]\ar@{-}[ddrr]& \color{blue}\CIRCLE\ar@{-}[dr]\ar@{-}[ddr]\ar@{-}[dll]\ar@{-}[ddll]&\\
\CIRCLE &&& \CIRCLE\\
\CIRCLE&&& \CIRCLE\\
& \color{blue}\CIRCLE\ar@{-}[r]\ar@{-}[ul]\ar@{-}[uul]\ar@{-}[uurr]\ar@{-}[urr]& \color{blue}\CIRCLE\ar@{-}[ur]\ar@{-}[uur]\ar@{-}[uull]\ar@{-}[ull]&\\
}$ }
\subfigure[] 
{
    \label{fig88c:sub:b}
$\xymatrix@M=0pt@W=0pt@R=15.78pt@C=15.78pt
{
& \color{blue}\CIRCLE\ar@{-}[r]\ar@{-}[dl]\ar@{-}[ddl]\ar@{-}[drr]\ar@{-}[ddrr]& \color{blue}\CIRCLE\ar@{-}[dll]\ar@{-}[dr]\ar@{-}[ddr]&\\
\CIRCLE &&& \CIRCLE\\
\CIRCLE&&& \CIRCLE\\
& \color{blue}\CIRCLE\ar@{-}[r]\ar@{-}[ul]\ar@{-}[uul]\ar@{-}[uurr]\ar@{-}[urr]& \color{blue}\CIRCLE\ar@{-}[uuu]\ar@{-}[ur]\ar@{-}[uur]\ar@{-}[ull]&\\
}$ }
\subfigure[] 
{
    \label{fig88c:sub:c}
$\xymatrix@M=0pt@W=0pt@R=15.78pt@C=15.78pt
{
& \color{blue}\CIRCLE\ar@{-}[r]\ar@{-}[dl]\ar@{-}[ddl]\ar@{-}[drr]\ar@{-}[ddrr]& \color{blue}\CIRCLE\ar@{-}[ddll]\ar@{-}[dr]\ar@{-}[ddr]&\\
\CIRCLE &&& \CIRCLE\\
\CIRCLE&&& \CIRCLE\\
& \color{blue}\CIRCLE\ar@{-}[r]\ar@{-}[ul]\ar@{-}[uul]\ar@{-}[uurr]\ar@{-}[urr]& \color{blue}\CIRCLE\ar@{-}[uuu]\ar@{-}[ur]\ar@{-}[uur]\ar@{-}[ull]&\\
}$ } \caption{Finding possible topology of $(8,8)$ absorption sets -- Step 4.}
\label{fig88c:sub} 
\end{figure}

\begin{figure}[!t]
\centering \subfigure[] 
{
    \label{fig12:sub:a}
$\xymatrix@M=0pt@W=0pt@R=27pt@C=10pt
{
\CIRCLE\ar@{-}[rr]\ar@{-}[drrrr]\ar@{-}[drr]\ar@{-}[ddr]\ar@{-}[d]&& \CIRCLE\ar@{-}[rr]\ar@{-}[drr]\ar@{-}[dll]\ar@{-}[d]&& \CIRCLE\ar@{-}[dllll]\ar@{-}[dll]\ar@{-}[ddl]\ar@{-}[d]\\
\CIRCLE\ar@{-}[drrr]\ar@{-}[dr]&& \CIRCLE\ar@{-}[dl]\ar@{-}[dr]&& \CIRCLE\ar@{-}[dlll]\ar@{-}[dl]\\
& \CIRCLE\ar@{-}[rr]&& \CIRCLE&\\
}$ }
\subfigure[] 
{
    \label{fig12:sub:b}
$\xymatrix@M=0pt@W=0pt@R=15.78pt@C=15.78pt
{
& \CIRCLE\ar@{-}[r]\ar@{-}[dl]\ar@{-}[ddl]\ar@{-}[drr]\ar@{-}[ddrr]& \CIRCLE \ar@{-}[dll]\ar@{-}[ddll]\ar@{-}[dr]\ar@{-}[ddr]&\\
\CIRCLE \ar@{-}[d]\ar@{-}[ddr]\ar@{-}[ddrr]&&& \CIRCLE\ar@{-}[d]\ar@{-}[ddl]\ar@{-}[ddll]\\
\CIRCLE\ar@{-}[dr]\ar@{-}[drr]&&& \CIRCLE\ar@{-}[dl]\ar@{-}[dll]\\
& \CIRCLE\ar@{-}[r]& \CIRCLE&\\
}$ }
\subfigure[] 
{
    \label{fig12:sub:c}
$\xymatrix@M=0pt@W=0pt@R=15.78pt@C=15.78pt
{
& \CIRCLE\ar@{-}[r]\ar@{-}[dl]\ar@{-}[ddl]\ar@{-}[drr]\ar@{-}[ddd]& \CIRCLE\ar@{-}[dll]\ar@{-}[dr]\ar@{-}[ddr]&\\
\CIRCLE \ar@{-}[ddrr]\ar@{-}[d]&&& \CIRCLE\ar@{-}[d]\\
\CIRCLE \ar@{-}[rrr]&&& \CIRCLE\\
& \CIRCLE\ar@{-}[ul]\ar@{-}[uul]\ar@{-}[uurr]\ar@{-}[urr]& \CIRCLE\ar@{-}[uuu]\ar@{-}[ur]\ar@{-}[uur]\ar@{-}[ull]&\\
}$ } \caption{Possible topologies of $(8,8)$ absorption
sets.}
\label{fig12:sub} 
\end{figure}
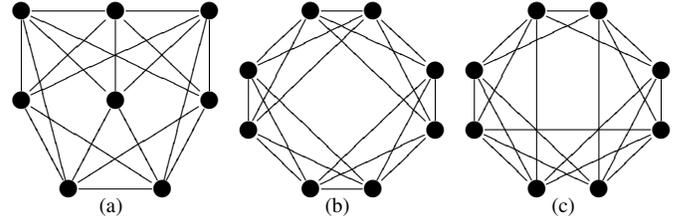

As Step 3, now there are two choices for the bottom-right node as shown in \figurename \ref{fig88b:sub:b}--\ref{fig88b:sub:c}:
\begin{enumerate}\setlength{\itemsep}{0pt}
\item \figurename \ref{fig88b:sub:b}: as Step 4, the top-right node only has one choice as shown in \figurename \ref{fig88c:sub:a}. Then the other nodes have no choice. We obtain another possible topology as shown in \figurename \ref{fig12:sub:b}.
\item \figurename \ref{fig88b:sub:c}: as Step 4, this top-right node has two choices:
\begin{enumerate}\setlength{\itemsep}{0pt}
\item \figurename \ref{fig88c:sub:b}: By connecting the remaining nodes, we obtain \figurename \ref{fig12:sub:c}.
\item \figurename \ref{fig88c:sub:c}: After connecting the remaining node, this gives us \figurename \ref{fig12:sub:a} again.
\end{enumerate}
\end{enumerate}

So eventually we obtain the other three
possible $(8,8)$ topologies as shown in \figurename \ref{fig12:sub}.

\section{Less Dominant Absorption Sets}\label{less}

We list some examples in this section to show the existence of less dominant absorption sets.

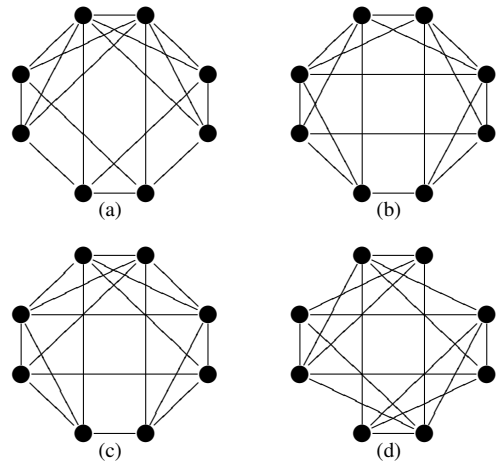
\begin{figure}[!t]
\centering \subfigure[] 
{
    \label{top812:sub:a}
$\xymatrix@M=0pt@W=0pt@R=15.78pt@C=15.78pt {
& \CIRCLE\ar@{-}[r]\ar@{-}[dl]\ar@{-}[ddl]\ar@{-}[drr]\ar@{-}[ddrr]\ar@{-}[ddd]& \CIRCLE \ar@{-}[dll]\ar@{-}[ddll]\ar@{-}[dr]\ar@{-}[ddr]\ar@{-}[ddd]&\\
\CIRCLE \ar@{-}[d]\ar@{-}[ddrr]&&& \CIRCLE\ar@{-}[d]\ar@{-}[ddll]\\
\CIRCLE\ar@{-}[dr]&&& \CIRCLE\ar@{-}[dl]\\
& \CIRCLE\ar@{-}[r]& \CIRCLE&\\
}$ }\qquad
\subfigure[] 
{
    \label{top812:sub:sub:b}
$\xymatrix@M=0pt@W=0pt@R=15.78pt@C=15.78pt {
& \CIRCLE\ar@{-}[r]\ar@{-}[dl]\ar@{-}[ddl]\ar@{-}[drr]\ar@{-}[ddd]& \CIRCLE \ar@{-}[dll]\ar@{-}[ddd]\ar@{-}[dr]\ar@{-}[ddr]&\\
\CIRCLE \ar@{-}[d]\ar@{-}[ddr]\ar@{-}[rrr]&&& \CIRCLE\ar@{-}[d]\ar@{-}[ddl]\\
\CIRCLE\ar@{-}[dr]\ar@{-}[rrr]&&& \CIRCLE\ar@{-}[dl]\\
& \CIRCLE\ar@{-}[r]& \CIRCLE&\\
}$ }\\
\subfigure[] 
{
    \label{top812:sub:sub:c}
$\xymatrix@M=0pt@W=0pt@R=15.78pt@C=15.78pt {
& \CIRCLE\ar@{-}[r]\ar@{-}[dl]\ar@{-}[ddrr]\ar@{-}[drr]\ar@{-}[ddd]& \CIRCLE \ar@{-}[dll]\ar@{-}[ddd]\ar@{-}[dr]\ar@{-}[ddll]&\\
\CIRCLE \ar@{-}[d]\ar@{-}[ddr]\ar@{-}[rrr]&&& \CIRCLE\ar@{-}[d]\ar@{-}[ddl]\\
\CIRCLE\ar@{-}[dr]\ar@{-}[rrr]&&& \CIRCLE\ar@{-}[dl]\\
& \CIRCLE\ar@{-}[r]& \CIRCLE&\\
}$ }\qquad
\subfigure[] 
{
    \label{top812:sub:sub:d}
$\xymatrix@M=0pt@W=0pt@R=15.78pt@C=15.78pt {
& \CIRCLE\ar@{-}[r]\ar@{-}[ddl]\ar@{-}[ddrr]\ar@{-}[drr]\ar@{-}[ddd]& \CIRCLE \ar@{-}[dll]\ar@{-}[ddd]\ar@{-}[ddll]&\\
\CIRCLE \ar@{-}[d]\ar@{-}[ddrr]\ar@{-}[rrr]&&& \CIRCLE\ar@{-}[d]\ar@{-}[ddl]\ar@{-}[ddll]\\
\CIRCLE\ar@{-}[drr]\ar@{-}[rrr]&&& \CIRCLE\ar@{-}[dll]\\
& \CIRCLE\ar@{-}[r]& \CIRCLE&\\
}$ } \caption{Some topologies of $(8,12)$ absorption sets.}
\label{top812:sub} 
\end{figure}

There are two classes of $(8,12)$ absorption sets:
\begin{enumerate}\setlength{\itemsep}{0pt}
\item $[6,6,4,4,4,4,4,4]$: $11,008$ such sets;
\item $[5,5,5,5,4,4,4,4]$: $33,408$ such sets.
\end{enumerate}
Some topologies are shown in \figurename \ref{top812:sub}.

\begin{table}[!t]
\renewcommand{\arraystretch}{1.1}
\caption{An example of $(8,14)$ absorption sets.}
\centering
\begin{tabular}{|c||c|c|c|c|c|c|}\hline
\bfseries Variable Nodes & \multicolumn{6}{c|}{\bfseries Six Neighboring Check Nodes} \\\hline\hline
0&56&120&184&248&312&376\\\hline
109&56&79&174&199&300&349\\\hline
1084&2&120&174&243&301&377\\\hline
116&46&76&184&243&256&372\\\hline
561&39&112&134&248&303&334\\\hline
870&2&76&135&192&264&334\\\hline
1091&46&79&135&237&303&329\\\hline
1970&0&69&134&199&264&329\\\hline
\end{tabular}
\end{table}

\begin{table}[!t]
\renewcommand{\arraystretch}{1.1}
\caption{An example of $(8,16)$ absorption sets.}
\centering
\begin{tabular}{|c||c|c|c|c|c|c|}\hline
\bfseries Variable Nodes & \multicolumn{6}{c|}{\bfseries Six Neighboring Check Nodes} \\\hline\hline
0&56&120&184&248&312&376\\\hline
109&56&79&174&199&300&349\\\hline
90&15&120&140&253&305&338\\\hline
1045&23&110&184&199&306&336\\\hline
1440&23&76&174&248&263&370\\\hline
39&15&79&143&207&271&335\\\hline
1048&19&76&143&253&262&354\\\hline
1444&19&110&140&207&317&326\\\hline
\end{tabular}
\end{table}

\begin{table}[!t]
\renewcommand{\arraystretch}{1.1}
\caption{An example of $(9,12)$ absorption sets.}
\centering
\begin{tabular}{|c||c|c|c|c|c|c|}\hline
\bfseries Variable Nodes & \multicolumn{6}{c|}{\bfseries Six Neighboring Check Nodes} \\\hline\hline
0&56&120&184&248&312&376\\\hline
109&56&79&174&199&300&349\\\hline
1563&29&120&150&217&264&336\\\hline
1628&40&75&171&248&264&373\\\hline
176&43&78&174&251&267&376\\\hline
314&8&125&150&251&300&368\\\hline
560&40&78&177&199&313&368\\\hline
1258&29&79&137&213&313&370\\\hline
1988&30&125&171&213&267&336\\\hline
\end{tabular}
\end{table}

\begin{table}[!t]
\renewcommand{\arraystretch}{1.1}
\caption{An example of $(9,16)$ absorption sets.}
\centering
\begin{tabular}{|c||c|c|c|c|c|c|}\hline
\bfseries Variable Nodes & \multicolumn{6}{c|}{\bfseries Six Neighboring Check Nodes} \\\hline\hline
0&56&120&184&248&312&376\\\hline
109&56&79&174&199&300&349\\\hline
90&15&120&140&253&305&338\\\hline
1460&2&79&184&238&307&365\\\hline
580&7&102&148&253&307&376\\\hline
890&15&87&181&229&261&349\\\hline
1194&38&87&148&228&319&360\\\hline
1775&2&126&176&228&261&338\\\hline
1881&33&84&176&229&300&360\\\hline
\end{tabular}
\end{table}

\begin{table}[!t]
\renewcommand{\arraystretch}{1.1}
\caption{An example of $(9,18)$ absorption sets.}
\centering
\begin{tabular}{|c||c|c|c|c|c|c|}\hline
\bfseries Variable Nodes & \multicolumn{6}{c|}{\bfseries Six Neighboring Check Nodes} \\\hline\hline
0&56&120&184&248&312&376\\\hline
109&56&79&174&199&300&349\\\hline
90&15&120&140&253&305&338\\\hline
629&15&86&161&248&268&381\\\hline
1063&1&109&178&236&312&349\\\hline
504&20&85&174&197&258&338\\\hline
802&49&86&154&197&300&363\\\hline
1299&6&124&178&247&305&381\\\hline
1789&20&109&150&247&265&363\\\hline
\end{tabular}
\end{table}

There exist two classes of $(10,10)$ absorption sets:
\begin{enumerate}\setlength{\itemsep}{0pt}
\item $[5,5,5,5,5,5,5,5,5,5]$: $192$ such sets;
\item $[6,6,6,5,5,5,5,4,4,4]$: unknown.
\end{enumerate}

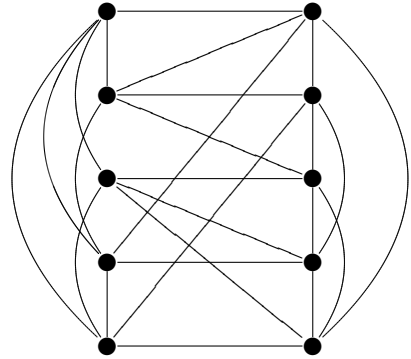
\begin{figure}[!t]
\centering $\xymatrix@M=0pt@W=0pt@R=25pt@C=70pt {
\CIRCLE\ar@{-}[r]\ar@{-}[d]\ar@/_1pc/@{-}[dd]\ar@/_2pc/@{-}[ddd]\ar@/_3pc/@{-}[dddd]&\CIRCLE \ar@{-}[ld]\ar@{-}[lddd]\ar@{-}[d]\ar@/^3pc/@{-}[dddd]\\
\CIRCLE\ar@{-}[r]\ar@{-}[rd]\ar@/_1pc/@{-}[dd] &\CIRCLE \ar@{-}[d]\\
\CIRCLE\ar@{-}[r]\ar@{-}[rd]\ar@{-}[rdd]\ar@/_1pc/@{-}[dd]&\CIRCLE \\
\CIRCLE\ar@{-}[r]\ar@{-}[d]&\CIRCLE\ar@{-}[u]\ar@/_1pc/@{-}[uu] \\
\CIRCLE\ar@{-}[r]\ar@{-}[ruuu]&\CIRCLE\ar@{-}[u]\ar@/_1pc/@{-}[uu] \\
}$ \caption{Topology of $[5,5,5,5,5,5,5,5,5,5]$ absorptions
sets.}\label{fig14}
\end{figure}

The average multiplicity of each variable node appeared in class $[5,5,5,5,5,5,5,5,5,5]$
is ${192\times 10}/{2048}=0.9375$. Once again, certain groups of
variable nodes share the same multiplicity, as listed in Table
\ref{table3}. As we can see, some groups are not involved at all.
Therefore, the average multiplicity of each involved variable node
in such sets is ${192\times 10}/{1408}\approx1.3636$. Like the $(8,8)$ ones,
for this class $\mu_{\rm max}=4$ and $\mathbf{v}_{\rm max}$ is an all-$1$ vector, since $\mathbf{V C}$ is a
probability matrix as well.

\begin{table}[!t]
\renewcommand{\arraystretch}{1.1}
 \caption{The
multiplicity of each variable node in class $[5,5,5,5,5,5,5,5,5,5]$ of $(10,10)$ absorption sets.}
\label{table3}
\centering
\begin{tabular}{|c|c||c|c|}\hline
\bfseries Variable Nodes & \bfseries Multiplicities & \bfseries Variable Nodes &
\bfseries Multiplicities\\\hline\hline
$0$---$63$&$1$&$1024$---$1087$&$1$\\\hline
$64$---$127$&$1$&$1088$---$1151$&$0$\\\hline
$128$---$191$&$1$&$1152$---$1215$&$0$\\\hline
$192$---$255$&$1$&$1216$---$1279$&$2$\\\hline
$256$---$319$&$0$&$1280$---$1343$&$2$\\\hline
$320$---$383$&$2$&$1344$---$1407$&$0$\\\hline
$384$---$447$&$0$&$1408$---$1471$&$1$\\\hline
$448$---$511$&$2$&$1472$---$1535$&$1$\\\hline
$512$---$575$&$1$&$1536$---$1599$&$1$\\\hline
$576$---$639$&$1$&$1600$---$1663$&$0$\\\hline
$640$---$703$&$1$&$1664$---$1727$&$2$\\\hline
$704$---$767$&$2$&$1728$---$1791$&$2$\\\hline
$768$---$831$&$0$&$1792$---$1855$&$1$\\\hline
$832$---$895$&$2$&$1856$---$1919$&$1$\\\hline
$896$---$959$&$1$&$1920$---$1983$&$0$\\\hline
$960$---$1023$&$0$&$1984$---$2047$&$0$\\\hline
\end{tabular}
\end{table}

Table \ref{table1010b} shows the existence of another class of $(10,10)$ absorption sets: $[6,6,6,5,5,5,5,4,4,4]$.

\begin{table}[!t]
\renewcommand{\arraystretch}{1.1}
\caption{Another class of $(10,10)$ absorption sets.}
\label{table1010b} \centering
\begin{tabular}{|c||c|c|c|c|c|c|}\hline
\bfseries Variable Nodes & \multicolumn{6}{c|}{\bfseries Six Neighboring Check Nodes} \\\hline\hline
0&56&120&184&248&312&376\\\hline
591&56&65&159&207&302&327\\ \hline
1405&32&120&159&225&314&337\\ \hline
1904&0&119&184&249&314&379\\ \hline
210&42&65&160&248&286&335\\ \hline
732&30&118&157&223&312&335\\ \hline
1676&36&77&183&223&286&376\\ \hline
616&30&119&160&194&275&373\\ \hline
834&42&85&157&249&302&373\\ \hline
892&36&118&142&225&275&327\\ \hline
\end{tabular}
\end{table}

\begin{table}[!t]
\renewcommand{\arraystretch}{1.1}
\caption{An example of $(10,12)$ absorption sets.}
\centering
\begin{tabular}{|c||c|c|c|c|c|c|}\hline
\bfseries Variable Nodes & \multicolumn{6}{c|}{\bfseries Six Neighboring Check Nodes} \\\hline\hline
0&56&120&184&248&312&376\\\hline
109&56&79&174&199&300&349\\\hline
90&15&120&140&253&305&338\\\hline
1460&2&79&184&238&307&365\\\hline
1320&46&67&140&248&307&320\\ \hline
1543&51&72&145&253&312&320\\ \hline
931&45&88&160&252&305&376\\\hline
9&46&110&174&238&302&366\\\hline
1104&33&67&145&252&282&349\\ \hline
1316&51&88&156&199&302&365\\\hline
\end{tabular}
\end{table}

\begin{table}[!t]
\renewcommand{\arraystretch}{1.1}
\caption{An example of $(10,14)$ absorption sets.}
\centering
\begin{tabular}{|c||c|c|c|c|c|c|}\hline
\bfseries Variable Nodes & \multicolumn{6}{c|}{\bfseries Six Neighboring Check Nodes} \\\hline\hline
0&56&120&184&248&312&376\\\hline
109&56&79&174&199&300&349\\\hline
90&15&120&140&253&305&338\\\hline
1460&2&79&184&238&307&365\\\hline
931&45&88&160&252&305&376\\\hline
9&46&110&174&238&302&366\\\hline
1121&15&110&156&198&315&321\\\hline
1316&51&88&156&199&302&365\\\hline
1432&33&121&160&226&315&338\\\hline
1549&45&121&142&198&300&366\\\hline
\end{tabular}
\end{table}

\begin{table}[!t]
\renewcommand{\arraystretch}{1.1}
\caption{An example of $(10,16)$ absorption sets.}
\centering
\begin{tabular}{|c||c|c|c|c|c|c|}\hline
\bfseries Variable Nodes & \multicolumn{6}{c|}{\bfseries Six Neighboring Check Nodes} \\\hline\hline
0&56&120&184&248&312&376\\\hline
109&56&79&174&199&300&349\\\hline
90&15&120&140&253&305&338\\\hline
1045&23&110&184&199&306&336\\\hline
176&43&78&174&251&267&376\\\hline
39&15&79&143&207&271&335\\\hline
305&23&78&132&201&259&335\\\hline
1048&19&76&143&253&262&354\\\hline
1189&43&76&132&234&300&326\\\hline
1444&19&110&140&207&317&326\\\hline
\end{tabular}
\end{table}

\begin{table}[!t]
\renewcommand{\arraystretch}{1.1}
\caption{An example of $(10,18)$ absorption sets.}
\centering
\begin{tabular}{|c||c|c|c|c|c|c|}\hline
\bfseries Variable Nodes & \multicolumn{6}{c|}{\bfseries Six Neighboring Check Nodes} \\\hline\hline
0&56&120&184&248&312&376\\\hline
109&56&79&174&199&300&349\\\hline
90&15&120&140&253&305&338\\\hline
170&49&124&184&196&283&366\\\hline
931&45&88&160&252&305&376\\\hline
253&63&124&174&226&259&336\\\hline
1121&15&110&156&198&315&321\\\hline
1432&33&121&160&226&315&338\\\hline
1549&45&121&142&198&300&366\\\hline
2016&9&113&156&196&259&349\\\hline
\end{tabular}
\end{table}

\begin{table}[!t]
\renewcommand{\arraystretch}{1.1}
\caption{An example of $(10,20)$ absorption sets.}
\centering
\begin{tabular}{|c||c|c|c|c|c|c|}\hline
\bfseries Variable Nodes & \multicolumn{6}{c|}{\bfseries Six Neighboring Check Nodes} \\\hline\hline
0&56&120&184&248&312&376\\\hline
109&56&79&174&199&300&349\\\hline
90&15&120&140&253&305&338\\\hline
358&52&68&184&255&315&327\\\hline
1056&10&115&135&248&300&333\\\hline
574&15&80&169&255&316&333\\\hline
712&52&125&147&198&316&347\\\hline
862&10&125&174&242&285&325\\\hline
1207&25&68&140&232&285&356\\\hline
1837&42&79&147&253&293&356\\\hline
\end{tabular}
\end{table}

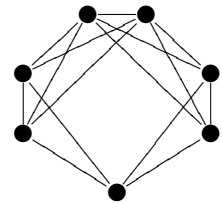
\begin{figure}[!t]
\centering $\xymatrix@M=0pt@W=0pt@R=15.78pt@C=3.3pt {
&& \CIRCLE\ar@{-}[rr]\ar@{-}[dll]\ar@{-}[ddll]\ar@{-}[drrrr]\ar@{-}[ddrrrr]&& \CIRCLE \ar@{-}[dllll]\ar@{-}[ddllll]\ar@{-}[drr]\ar@{-}[ddrr]&&\\
\CIRCLE \ar@{-}[d]\ar@{-}[ddrrr]&&&&&& \CIRCLE\ar@{-}[d]\ar@{-}[ddlll]\\
\CIRCLE\ar@{-}[drrr]&\hspace{10pt}&&&&\hspace{10pt}& \CIRCLE\ar@{-}[dlll]\\
&&& \CIRCLE&&&\\
}$ \caption{Topology of $(7,12)$ absorptions sets.}\label{top712}
\end{figure}

Note that $(7,12)$ and $(9,14)$ absorption sets (though not all of
them) can be obtained by removing one node from $(8,8)$ and
$(10,10)$ absorption sets, respectively. For example, there are
$179,648$ $(7,12)$ absorption sets. They all have the topology
\figurename \ref{top712}, which can be obtained from \figurename
\ref{fig13:sub} by removing one node. However, the $(8,8)$ sets
generate $14272\times8=114176$ $(7,12)$ absorption sets (no
duplicates). Hence there are $179648-114176=65472$ $(7,12)$ sets
that are not contained in the $(8,8)$ ones.

\ifCLASSOPTIONcaptionsoff
  \newpage
\fi



%



\bibliographystyle{IEEEtran}
\bibliography{IEEEabrv,ref}

%










\end{document}